\documentclass[aps,prb,reprint,superscriptaddress]{revtex4-2}

\usepackage{amsmath, newtxtext, newtxmath}
\usepackage{graphicx}
\usepackage{color}
\definecolor{bblue}{rgb}{0, 0.0, 0.8}
\definecolor{rred}{rgb}{0.7, 0.0, 0.0}

\usepackage{hyperref}
\hypersetup{
    colorlinks=true,        
    linkcolor=rred,          
    citecolor=rred,        
    filecolor=rred,      
    urlcolor=rred           
}
\newcommand{\lco}{LaCoO$_{3}$}

\begin{document}

\title{A review on magnetic field induced spin crossover in \lco{} up to 600 T}

\author{Akihiko~Ikeda}
\email[E-mail: ]{a-ikeda@uec.ac.jp}
\affiliation{University of Electro-Communications, Chofu, Tokyo 182-8585, Japan}
\author{Yasuhiro~H.~Matsuda}
\affiliation{Institute for Solid State Physics, University of Tokyo, Kashiwa, Chiba 277-8581, Japan}
\author{Keisuke~Sato}
\affiliation{National Institute of Technology, Ibaraki College, Hitachinaka, Ibaraki 312-0011, Japan}
\author{Joji~Nasu}
\affiliation{Department of Physics, Tohoku University, Sendai, Miyagi 980-8578, Japan}
\affiliation{PRESTO, Japan Science and Technology Agency, Honcho Kawaguchi, Saitama 332-0012, Japan}

\date{\today}

\begin{abstract}
\lco{} is known for its two-step spin crossover as a function of temperature.
Despite efforts spanning over half a century, the origin of this phenomenon is still debated, particularly regarding how the microscopic spin states are involved in the observed macroscopic two-step spin crossover.
High magnetic field studies on \lco{} are performed because the magnetic field-induced spin crossover is induced, where the magnetic excited states become more stable in high magnetic fields than the non-magnetic ground states.
This review focuses on the findings in \lco{} at high magnetic fields over the last decade. 
A complex phase diagram has been revealed at high magnetic fields instead of solving the conventional problem of \lco{}.
It suggests that appreciable spin state correlations are in play in \lco{}. The possibility of exciton condensation is also discussed.
\end{abstract}

\maketitle

\tableofcontents

\section{Introduction}

Transition metal oxide with Perovskite structure is a fertile ground for numerous many-body phenomena like ferroelectricity, metal-insulator transition, high-temperature superconductivity, multiferroicity, Mott transition, and magnetic transitions \cite{ImadaRMP1998, TokuraRPP2006}.
On the other hand, the spin crossover phenomenon is a change in the spin states of a single ion observed in coordination complexes.
Spin crossover is a single-ion physics generally well described by considering the energy level scheme of multi-electron states in a single ion.
The level scheme results from the competition between Hund's coupling and crystal field splitting at the transition metal ions surrounded by the ligands, forming high-spin and low-spin states.

Cobaltites are unique systems among transition metal oxides due to their spin state degree of freedom.
\lco{} is especially well-known for its two-step spin crossover with temperature evolution.
Unlike coordination complexes, ligands are shared in Co oxides.
It leads to the appreciable interaction between neighboring Co ions and their spin states.
This gives a basis for various spin state ordered phases in cobaltites.
This review focuses on the magnetic field effect.
Thus, please refer to a thorough review \cite{SCObook} for the long-standing controversy over the spin state evolution in \lco{}.

\begin{figure}
\begin{center}
\includegraphics[width = \columnwidth]{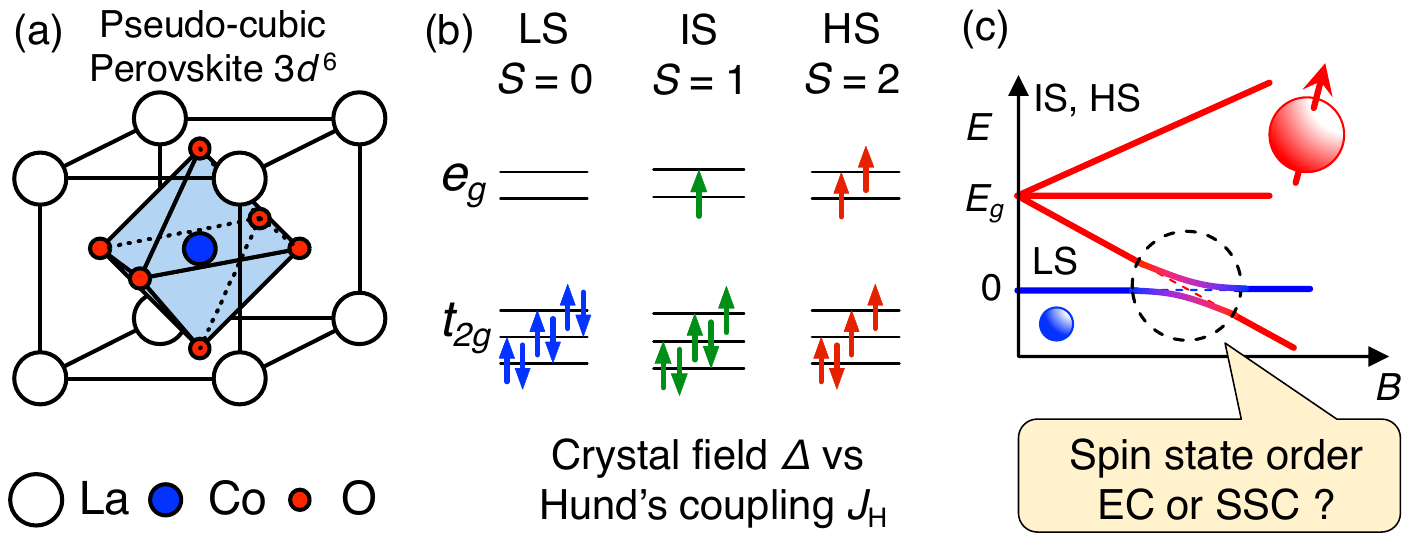}
\caption{A schematic drawing of (a) crystal structure, (b) spin states, and (c) magnetic field effect on \lco{}. 
\label{schem}
}
\end{center}
\end{figure}

\begin{figure}
\begin{center}
\includegraphics[width = 0.8\columnwidth]{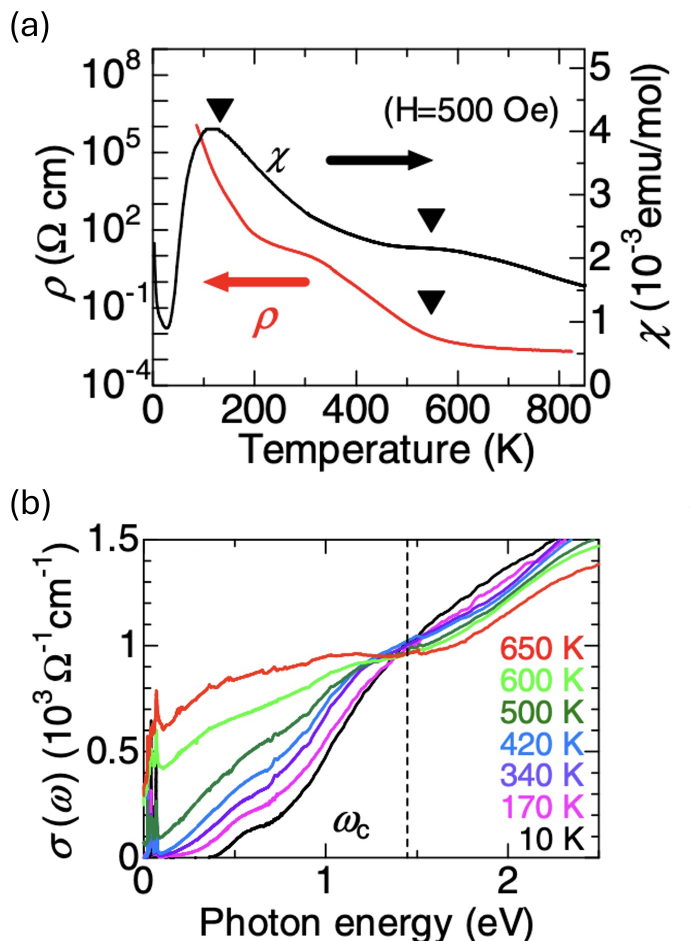}
\caption{(a) The magnetic susceptibility and electric conductivity of \lco{} as a function of temperature.
(b) The optical conductivity at various temperatures that ranges from 650 to 10 K. 
Reprinted from Ref. \onlinecite{DoiPRB2014} with a re-arrangement, $\copyright$ 2014 American Physical Society. \label{doi}}
\end{center}
\end{figure}

\begin{figure}
\begin{center}
\includegraphics[width = 0.8\columnwidth]{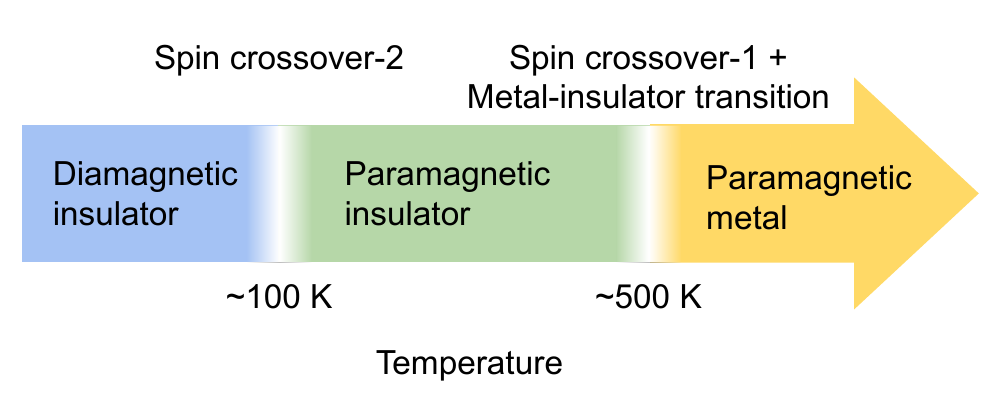}
\caption{The schematic drawing of the temperature evolution of the macroscopic properties of \lco{}. \label{sco}}
\end{center}
\end{figure} 

\lco{} shows the two-step spin crossover in the magnetic susceptibility and electric conductivity as functions of temperature \cite{DoiPRB2014} as shown in Fig. \ref{doi}.
The high-temperature phase above $\sim500$ K is a paramagnetic metal.
Upon decreasing temperature across $\sim500$ K, the electric conductivity becomes semiconducting, indicating the formation of an appreciable charge gap.
At the same time, the magnetic susceptibility sustains the Curie paramagnetic behavior with a reduction of the effective magnetic moment.
Thus, the intermediate temperature phase between $\sim100$ K and $\sim500$ K is a paramagnetic insulator.
With a further decrease in temperature, at $\sim100$ K, the magnetic susceptibility starts to decrease, losing the Curie paramagnetic moment.
The magnetic susceptibility becomes tiny below $\sim30$ K, indicating a spin-gap behavior where the ground state of \lco{} is a spin-singlet state with a magnetic excited state at $\sim$100 K above the ground state, that is, the spin-gap.
The calculated spin gap from the magnetic susceptibility is around 100 K.
The charge gap calculated from the electric conductivity is 140 meV \cite{EnglishPRB2002}.
The charge gap observed in the optical conductivity measurement is 300 meV [Fig. \ref{doi}].
The charge gap is larger than the spin gap by an order of magnitude.
The spin crossover is rather a robust phenomenon that even under a high magnetic field of 33 T, the magnetic susceptibility overlaps with the data at 10 T and 0.05 T \cite{HochPRB2009}.
The temperature evolution of the macroscopic properties of \lco{} is summarized in Fig. \ref{sco}.

\section{Previous experiments up to 100 T}

\begin{figure}
\begin{center}
\includegraphics[width = 0.7\columnwidth]{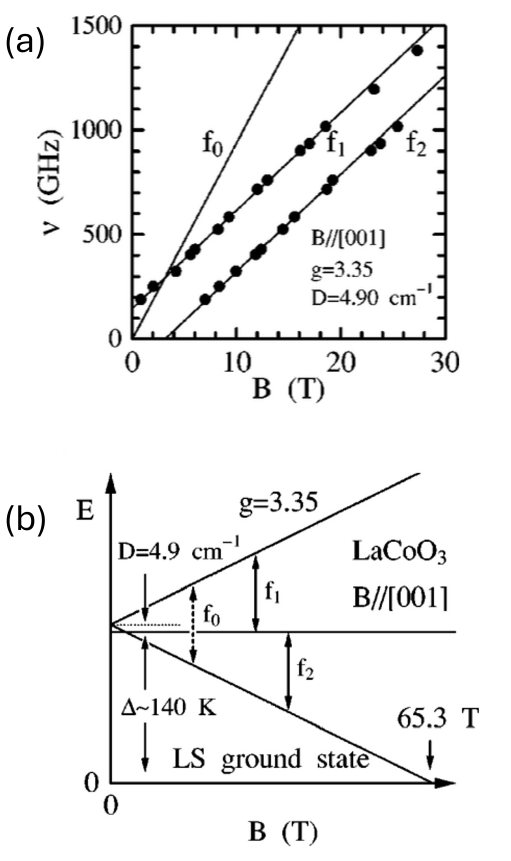}
\caption{(a) Electron spin resonance as a function of photon energy and applied magnetic fields.
(b) Obtained energy level scheme of \lco{}. 
Reprinted from Ref. \onlinecite{NoguchiPRB2002} with a re-arrangement, $\copyright$ 2002 American Physical Society.  \label{noguchi}}
\end{center}
\end{figure}

\begin{figure*}
\begin{center}
\includegraphics[width = \textwidth]{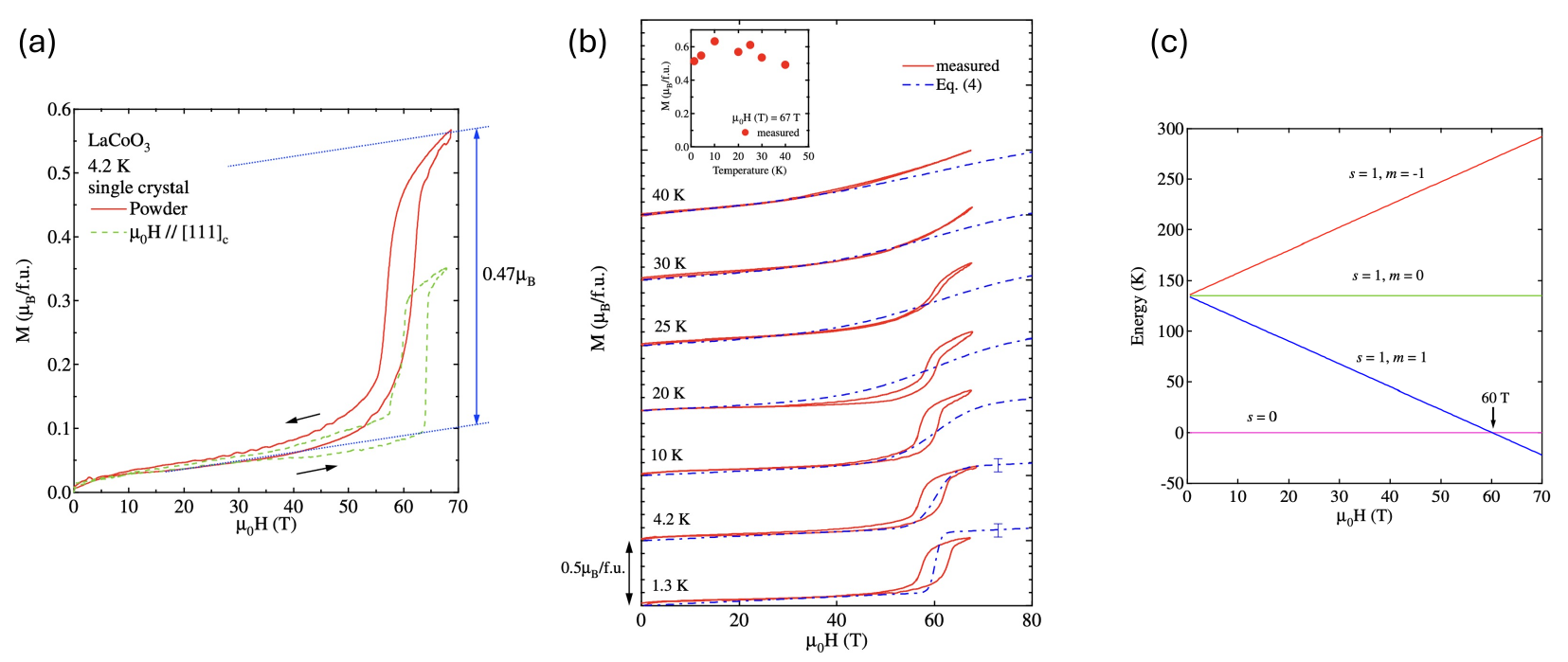}
\caption{(a) Result of a magnetization measurement up to a pulsed high magnetic field of 70 T.
(b) The temperature dependence of the high magnetic field induced spin state transition.
(c) The proposed energy diagram.
Reprinted from Ref. \onlinecite{SatoJPSJ2009} with a re-arrangement, $\copyright$ 2009 The Physical Society of Japan.  \label{sato}}
\end{center}
\end{figure*}

\begin{figure*}
\begin{center}
\includegraphics[width = \textwidth]{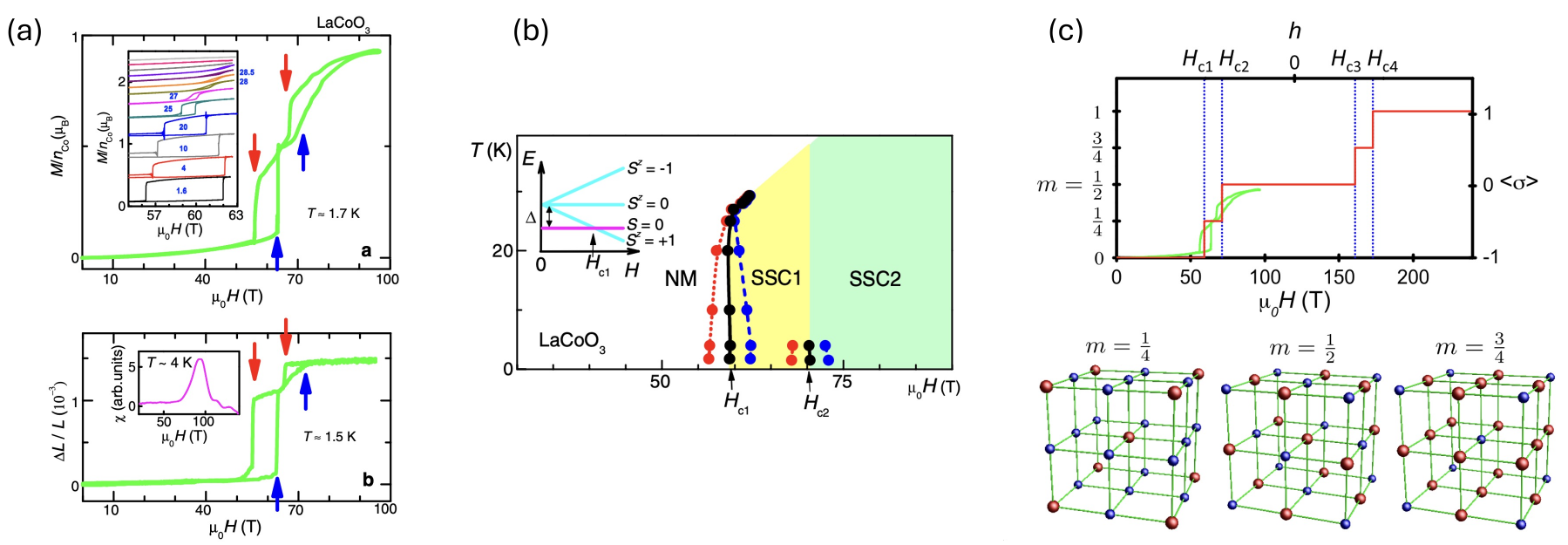}
\caption{(a) Result of magnetization measurement and magnetostriction measurement up to 100 T.
(b) The phase diagram up to 100 T and 30 K.
(c) The proposed magnetization evolution up to 200 T with schematic drawings of spin state crystals realized at each magnetization change.
Reprinted from Ref. \onlinecite{MoazPRL2012} with a re-arrangement, $\copyright$ 2012 American Physical Society. \label{moaz}}
\end{center}
\end{figure*}

\begin{figure}
\begin{center}
\includegraphics[width = 0.7\columnwidth]{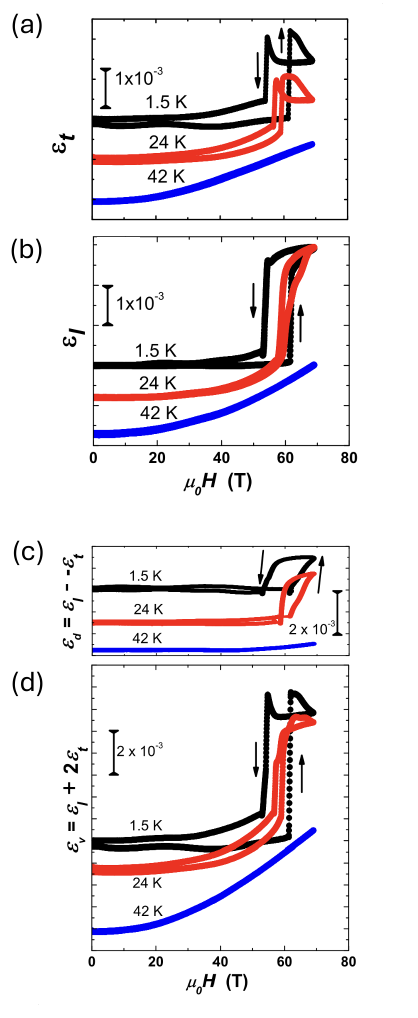}
\caption{(a) Results of transversal and (b) longitudinal magnetostriction measurement on \lco{} up to 70 T.
(c) The distortion and (d) volume expansion calculated from the longitudinal and transversal magnetostriction.
 Reprinted from Ref. \onlinecite{RotterSR2014} with a re-arrangement, $\copyright$ 2020 Licensed under CC BY 4.0. \label{rotter}}
\end{center}
\end{figure}

The first report on a pulsed high magnetic field study on \lco{} was the electron spin resonance study to investigate the controversy over the involvement of high-spin or intermediate spin states in the spin crossover at 100 K \cite{NoguchiPRB2002}.
They measured electron spin resonance experiment for a single crystalline \lco{} in various crystal orientations.
The single crystal was carefully de-twined to have only a single domain of the rhombohedrally distorted sample.
The results are summarized in Fig. \ref{noguchi}.
They observed absorption peaks as a function of magnetic fields with wavelength variations.
Their results are summarized in the schematic energy diagram in Fig. \ref{noguchi}.
The most important observation is that the excited state is a spin-triplet state with a $g$-factor of 3.35.
The crystal field and $g$-factor are isotropic with a slight anisotropy.
The interpretation is that the first excited state originates from the high-spin state with $S=2$ and $g=2$ and not the intermediate spin state with $S=1$ and $g=2$.
The high-spin state is modified with trigonal distortion and spin-orbit coupling, pushing the $S_{z} = \pm1$ further upwards.
As a result, the seemingly spin-triplet state is observed with an effective $g$-factor of 3.35.
This is theoretically supported by several multiplet calculations \cite{RopkaPRB2003} and a heat capacity measurement assuming such an energy diagram \cite{KyomenPRB2005}.
Strikingly, they predicted by the obtained energy diagram that the magnetic field-induced spin crossover will happen at 65.3 T as shown in Fig. \ref{noguchi}.

In 2009, Sato {\it et al.} discovered a sharp spin state transition at $\sim60$ T at 4.2 K \cite{SatoJPSJ2009}.
The results are shown in Fig. \ref{sato}.
The transition field of $\sim60$ T shows quite a good agreement with the prediction by Noguchi {\it et al.}\cite{NoguchiPRB2002}. 
However, several points are not in accordance with the simple energy diagram presented by Noguchi {\ et al.}.
The first feature is that the magnetization jump is only $\sim0.5$ $\mu_{\rm{B}}$/f.u.
It is much smaller than the expected saturation value of 2 $\mu_{\rm{B}}$/f.u. and 4 $\mu_{\rm{B}}$/f.u. for a high and intermediate spin state of Co$^{3+}$ with octahedral crystal field, respectively.
This indicates that \lco{} is far from its complete polarization at $\sim60$ T and that there should be more at a higher magnetic field region.
The second feature is that they clearly show the sharp transition in a first-order manner with hysteresis with the spin gap behavior below the transition magnetic field $\sim60$ T. 
This feature shows that the cooperative feature is present in \lco{}, which was not evident in the temperature-induced spin crossover.
The third feature is the temperature dependence of the spin state transition.
They monotonously become smeared with increasing temperature from 4. 2 K up to 40 K.
On the other hand, the transition magnetic field is not dependent on temperature, which does not agree with the simple energy level scheme picture where the transition field is expected to decrease with increasing temperature aided by the entropy effect.
This study by Sato {\it et al.} triggers all the later studies on \lco{} at high magnetic fields.

To extend the magnetic field range for search, Altarawneh {\it et al.} conducted a cooperative study up to 100 T employing magnetization measurement based on proximity detector oscillator \cite{GhannadzadehRSI2011}, magnetostriction measurement utilizing fiber Bragg grating (FBG) \cite{DaouRSI2010}, and the 100 T pulsed magnetic field with flywheel generator.
They successfully observed the second transition above 70 T that followed the known transition at 60 T \cite{MoazPRL2012} as shown in Fig. \ref{moaz}.
Here, they used a single crystalline sample.
They conducted a temperature dependence study, finding that the transition fields shift to higher magnetic fields with increasing temperatures beyond 25 K up to 30 K.
However, the transition magnetic field exceeds the available magnetic field of 100 T above 30 K.
The two-step transition at high magnetic fields motivated them to propose a classical Ising model with the first and the second nearest neighbor interactions.
The model predicts a possible two-step transition at 160 T and 170 T, where the saturation of the magnetic moment will happen at the latter transition, as shown in Fig. \ref{moaz}.

Rotter {\it et al.} further investigated the transition at 60 T using the FBG magnetostriction method by realizing the measurement of longitudinal ($\epsilon_{l} = \Delta L / L \parallel B$) and transversal ($\epsilon_{t} = \Delta L \perp B$) magnetostriction  \cite{RotterSR2014}.
They deduced the volume change $\epsilon_{v} = \epsilon_{l} + 2\epsilon_{t}$ and distortion $\epsilon_{d} = \epsilon_{l} - \epsilon_{t}$ using the simple relations.
The obtained results of $\epsilon_{l}$, $\epsilon_{t}$, $\epsilon_{v}$, and $\epsilon_{d}$ are displayed in Fig. \ref{rotter}.
The feature they found is that the transition is isotropic.
At the sharp transition at 60 T, only $\epsilon_{v}$ changes with little change in $\epsilon_{d}$.
At above 60 T, $\epsilon_{d}$ show changes with the change of $\epsilon_{v}$.
Based on this observation, they concluded that the transition is more driven by the high spin state than the intermediate spin state.
Based on the discussion, the intermediate spin state has the orbital degree of freedom in the $e_{g}$ orbital. 
In contrast, the high spin state has the orbital degree of freedom in $t_{2g}$, which should lead to the large anisotropy in the intermediate spin state, and the high spin state should be isotropic.

The experiments so far only uncover a small portion of high magnetic field phases in \lco{}.
Thus, a study above 100 T was needed.

\begin{figure*}
\begin{center}
\includegraphics[width = 0.8\textwidth]{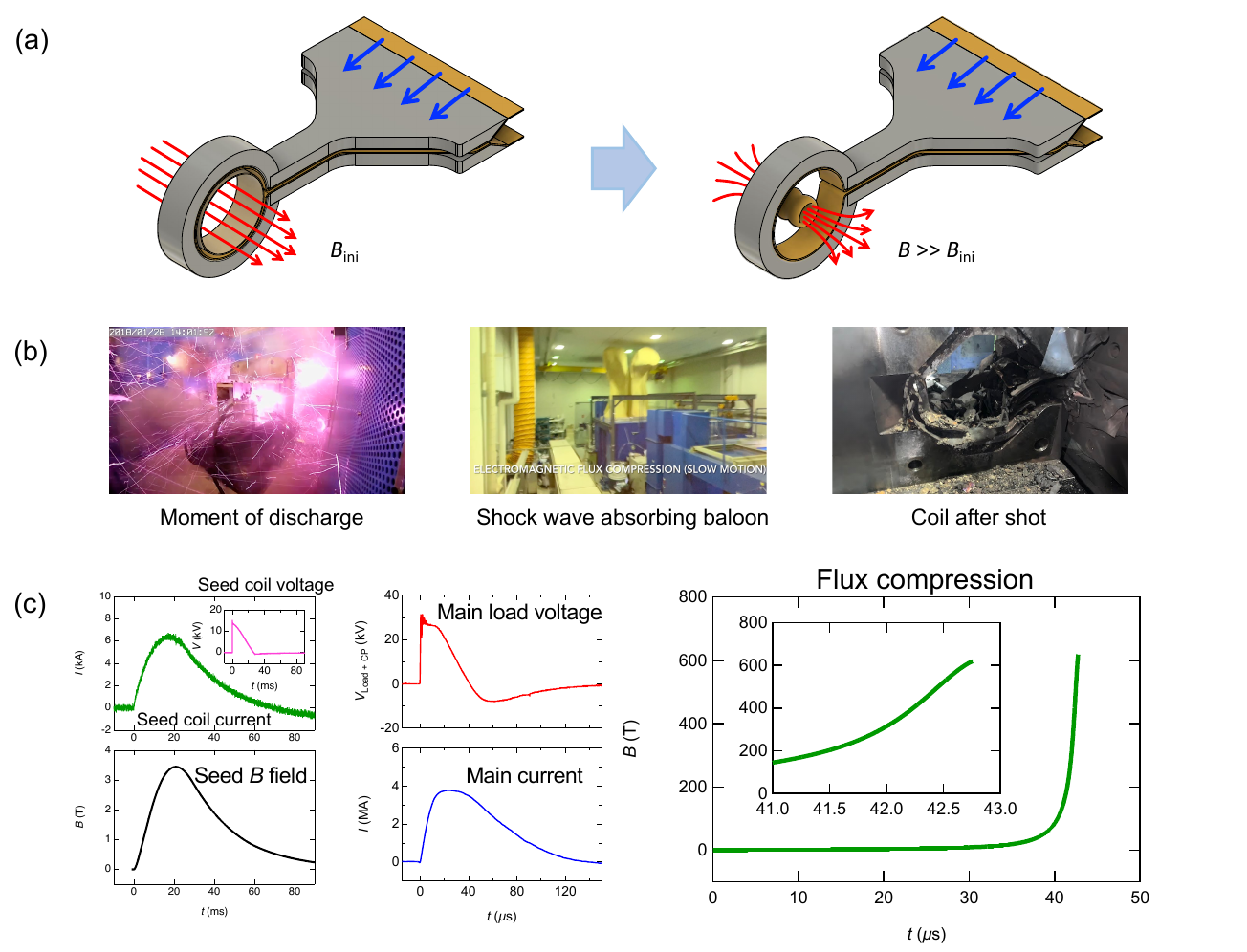}
\caption{(a) Schematic drawings, (b) photographs, (c) and waveforms of the generated magnetic field and current for the primary coil and seed coils, which are involved in the electromagnetic flux compression technique installed in the institute for Solid State Physics, the University of Tokyo, Japan in 2014.
Reprinted from Ref. \onlinecite{IkedaNC2023}, $\copyright$ 2020 Licensed under CC BY 4.0. \label{emfc}}
\end{center}
\end{figure*}

\begin{figure*}
\begin{center}
\includegraphics[width = 0.75\textwidth]{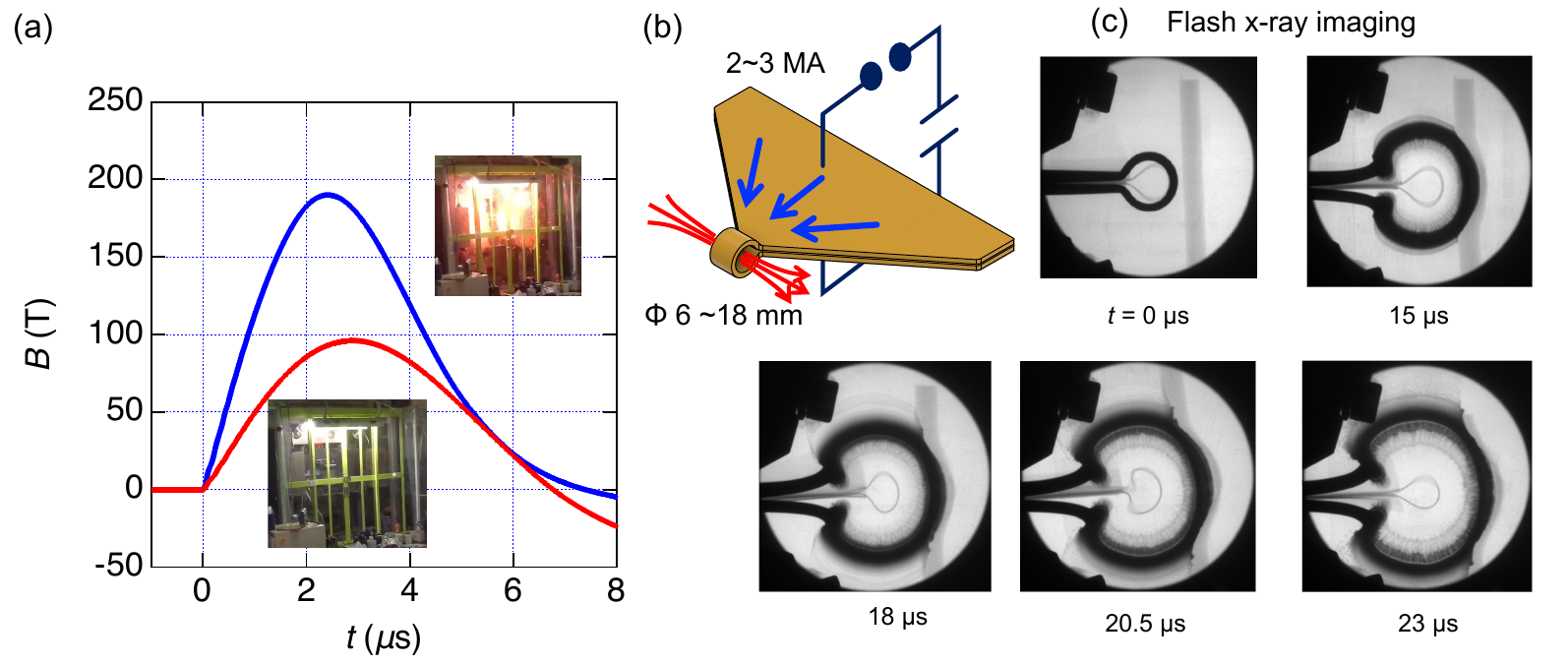}
\caption{(a) waveforms of generated magnetic fields, (b) schematic drawing, and (c) a series of flash x-ray radiography in the single turn coil method.\label{stc}}
\end{center}
\end{figure*}

\begin{figure*}
\begin{center}
\includegraphics[width = 0.6\textwidth]{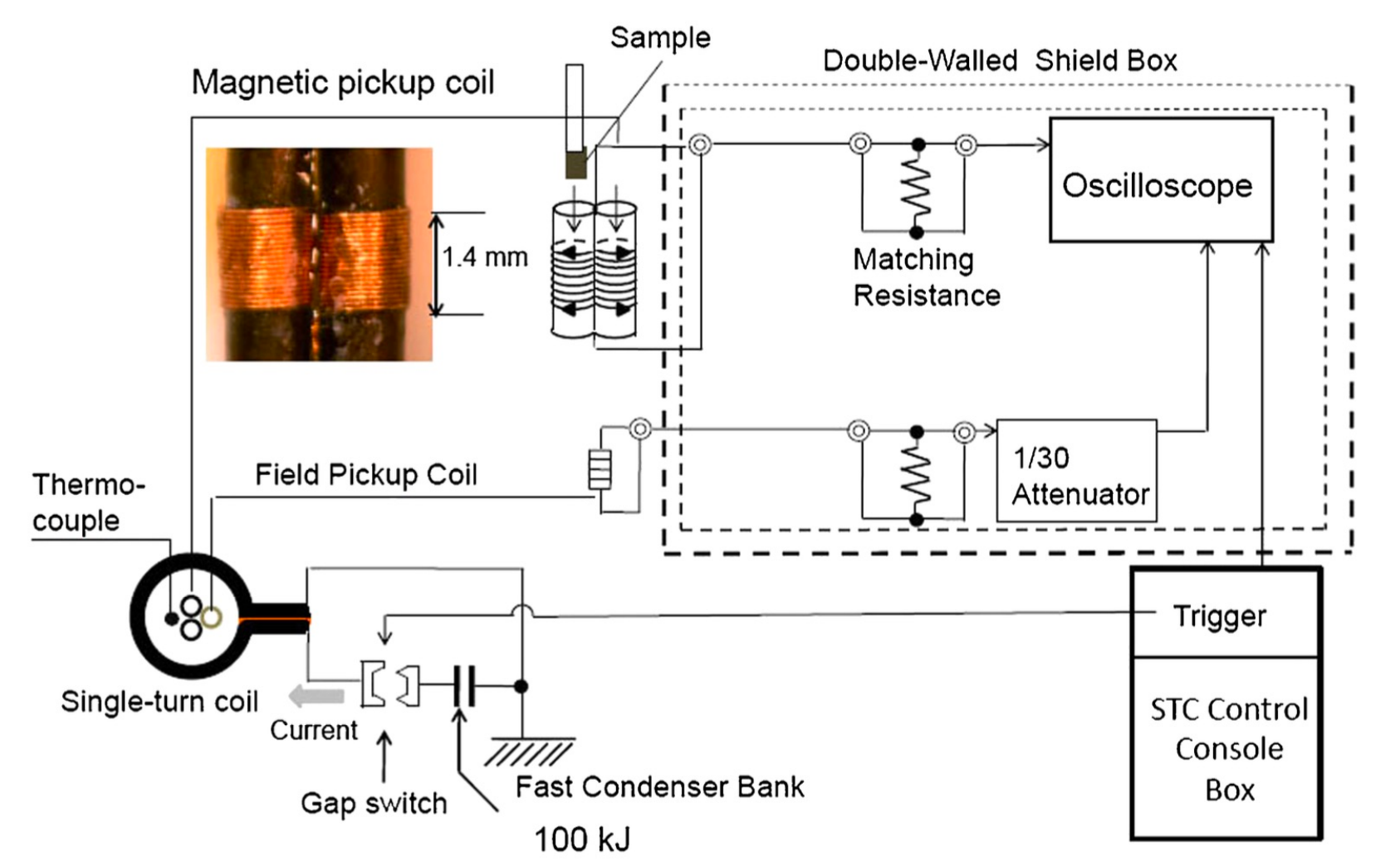}
\caption{Schematic drawing of magnetization measurements above 100 T using induction pick-up coil and single turn coil method. 
Reprinted from Ref. \onlinecite{TakeyamaJPSJ2012}, $\copyright$ 2012  The Physical Society of Japan. \label{mag}}
\end{center}
\end{figure*}

 \begin{figure*}
\begin{center}
\includegraphics[width = 0.8\textwidth]{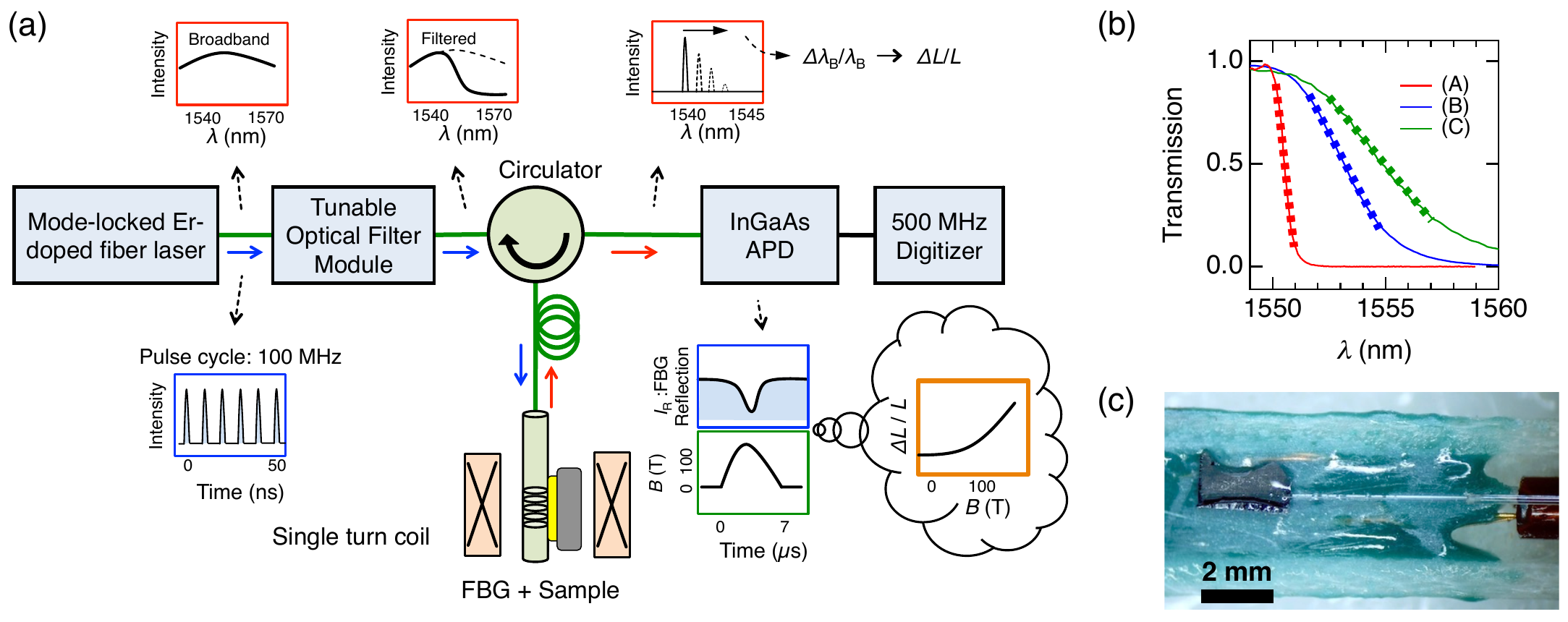}
\caption{(a) Schematic drawing of magnetostriction measurements above 100 T using FBG and optical filter method. 
The shift of the Bragg wavelength is converted to the change of the intensity of the reflected light from the FBG by utilizing the optical filter method.
The light intensity change is detected by an avalanche photodiode (APD) with 100 MHz high speed in a single shot measurement.
(b) Examples of the spectrum of optical filters used.
(c) A photograph of a solid sample glued to an optical fiber with FBG 
Reprinted from Ref. \onlinecite{IkedaRSI2017}, $\copyright$ 2017 American Institute of Physics. \label{fbg}}
\end{center}
\end{figure*}

\section{Experiments above 100 T}
\subsection{Methods}

One needs a destructive pulse magnet to obtain a high magnetic field well beyond 100 T experimentally.
Electromagnetic flux compression is a method to obtain a high magnetic field well beyond 500 T.
A new instrument of electromagnetic flux compression is installed in the Institute for Solid State Physics, the University of Tokyo, Kashiwa, Japan, where, in 2018, we reported the generation of the world's highest magnetic field indoor of 1200 T \cite{NakamuraRSI2018}.
The principle of electromagnetic flux compression is briefly explained.
First, a seed pulse magnet generates the seed magnetic field of $\sim3$ T inside a metallic cylinder of 11 cm diameter, a liner.
The seed magnetic field inside the liner is rapidly compressed to a diameter of less than 1 cm to obtain a few hundred times larger magnetic field.
This is realized using the implosion of the metallic cylinder, which is driven by the electromagnetic repulsive force generated by injecting the pulsed current of 2-5 MA to the primary coil.
The pulse width of the seed magnetic field and primary current are about 50 ms and 100 $\mu$s, respectively.
The final waveform of the generated magnetic field is shown in Fig. \ref{emfc}.
Right after the magnetic field is generated, a large explosion occurs due to the mechanical energy of the coils.
The samples, cryostat, and probes do not survive the explosion.
The pulse power for generating the primary current of 5 MA is rated at 50 kV and 5 MJ, which occupies the whole room of 900 m$^{2}$.
The explosion happens in the next room with a special chamber equipped with a shock-absorbing balloon.
In addition to the 5 MJ system, they installed the 2 MJ system in parallel to generate fields up to 400 T.

One can obtain a pulsed high magnetic field up to 250 T using single turn coil method in the same institute \cite{Miura2003}.
In the single turn coil method, a current of 2-3 MA is injected into a copper coil with a single turn.
The pulse duration is about 7 $\mu$s. 
The destruction of the coil follows the generation of the magnetic field.
The coils deform and scatter outwards.
Therefore, the sample, cryostat, and probes are kept safe after each magnetic field generation.
One can repeat the experiment by replacing the exploded single turn coil with a new one, which is a beauty of the single turn coil method compared to the electromagnetic flux compression method.

\begin{figure*}
\begin{center}
\includegraphics[width = \textwidth]{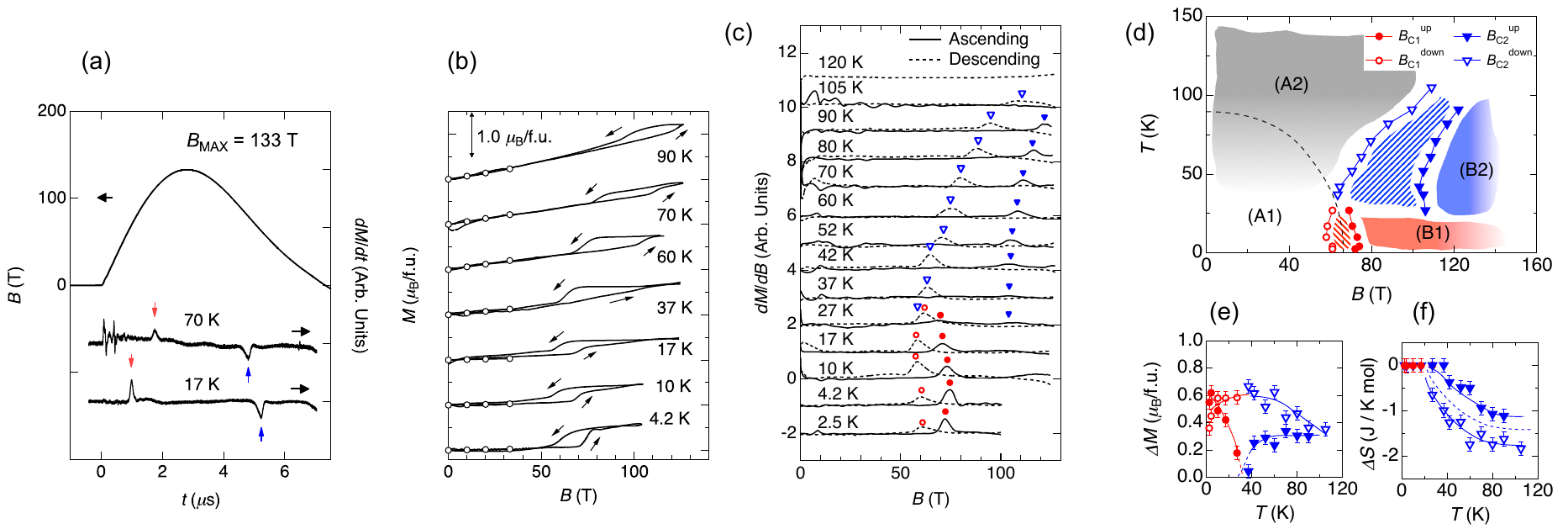}
\caption{Results of magnetization measurements on \lco{} up to 130 T using single turn coil method and induction method for the magnetization measurement. 
(a) $dM/dt$ as a function of time, (b) magnetization as a function of magnetic field,  (c) derivative of magnetization as a function of magnetic field, (d) temperature-magnetic field phase diagram, (e) the change of magnetization jump at the phase boundary, (f) the calculated change of entropy at the phase boundary using  Clausius-Clapeyron relation of the first order phase transition.
Reprinted from Ref. \onlinecite{IkedaPRB2016LCO} with a re-arrangement, $\copyright$ 2016 American Physical Society. \label{mag100t}}
\end{center}
\end{figure*}

Magnetization measurement above 100 T is performed using an induction method \cite{TakeyamaJPSJ2012}, which is schematically presented in Fig. \ref{mag}.
The magentization is measured as the induction signal from the magnetization pick up coil which is made of a pair of pick-up coils with a diameter of 1 mm placed in parallel.
Using this pick-up coil, the induction voltage from only the magnetization change of the sample inserted either one of the pick-up coil is measured.
The induction voltage from the external magnetic field is canceled out because the pair of the pick-up coils are connected in that way.
Still, however, to subtract the non-negligible background contribution, we have to measure two successive data with and without the sample by generating magnetic field pulse two times.
This fact that the induction measurement is not a single shot measurement is a considerable drawback of the technique because the 100 T generation relies on non-linear phenomena of air-gap discharge which is not easy to reproduce.
The technique is successful in single turn coil method.
However, it is not readily applicable to the electromagentic flux compression technique.
One reason is that the reproducibility of the magnetic field is poor.
Another reason is that one can not use the identical pick-up coil because it is lost in each magnetic field generation in the electromagnetic flux compression.
This aspect limits our measurement of magnetization up to 200 T or so.
Further innovation in magnetization measurement is mandatory for its realization at 1000 T.

\begin{figure*}
\begin{center}
\includegraphics[width = \textwidth]{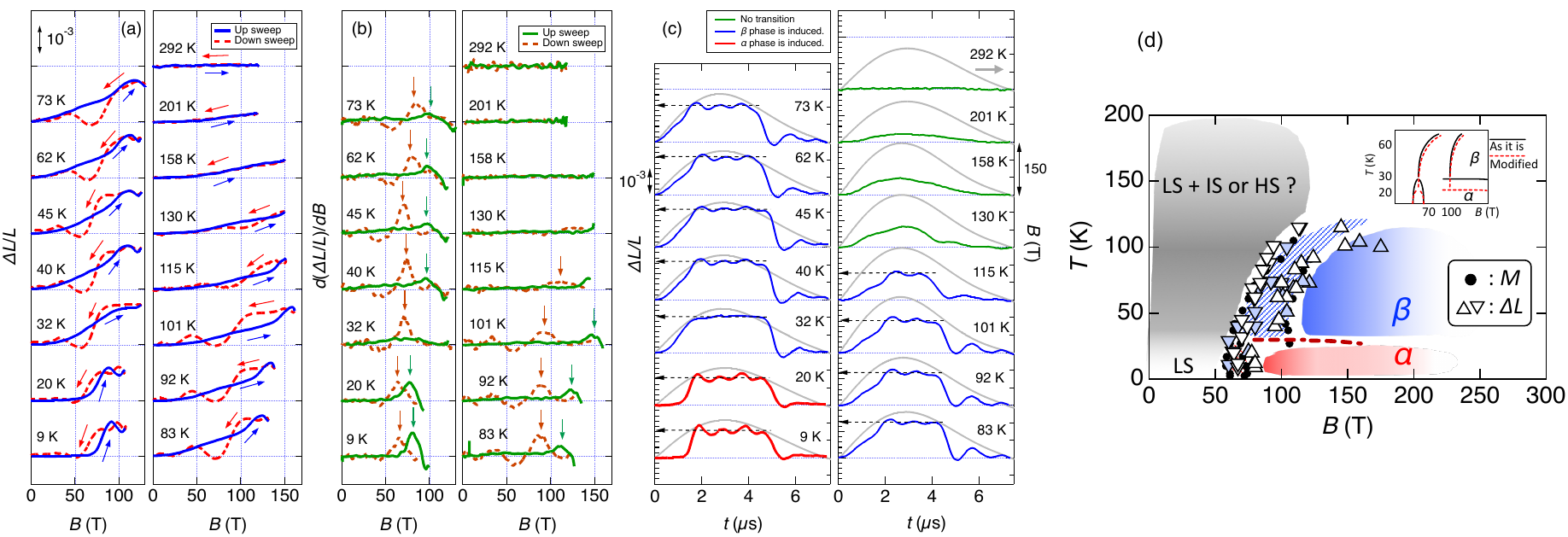}
\caption{Results of magnetostriction measurement up to 150 T on \lco{} using FBG and optical filter method. 
(a) Magnetostriction as a function of magnetic field, 
(b) Derivative of magnetostriction as a function of magnetic field, 
(c) Magnetostriction and magnetic field as a function of time, 
(b) A phase diagram summarizing the magnetostriction and magnetization data. 
Reprinted from Ref. \onlinecite{IkedaPRL2020} with a re-arrangement, $\copyright$ 2020 Licensed under CC BY 4.0. \label{fbg01}}
\end{center}
\end{figure*}

Magnetostriction above 100 T is performed using the FBG-based optical filter method, which we developed \cite{IkedaRSI2017}.
The method is a single-shot measurement.
Thus, it applies to the 100 T experiment in the single turn coil method and the 1000 T experiment in the electromagnetic flux compression experiment.
FBG is an optical fiber with some region of a few millimeters where modulation of the refractive index is present at the core.
The modulation works as the Bragg grating, reflecting only the light with the Bragg wavelength backward with close to 100 \% reflectivity.
Upon the physical elongation or shrinkage of the FBG itself, the Bragg wavelength shifts.
Hence, by optically measuring the Bragg wavelength, one can measure the length change of the optical fiber and the sample glued to the optical fiber.
Daou {\it et al.} first reported the magnetostriction measurement using FBG with pulse magnets \cite{DaouRSI2010}.
The obtained resolution is about 10$^{-7}$.
It was developed for the measurement with milliseconds pulsed magnetic fields with a time resolution of 47 kHz.
After the report, it is employed for various measurements up to 100 T \cite{JaimePNAS2012, JaimeSensors2017}.
There was a technical difficulty in applying the FBG magnetostriction method to above 100 T region because the destructively generated magnetic fields last only a few $\mu$ seconds.
The measurement speed was limited by the detection scheme of the wavelength of the reflected light from fiber Bragg gratings.
The previous method relies on fast optical spectroscopy with 50 kHz.
However, we needed 100 MHz measurements for the 100 T and the 1000 T experiments.
So, we devised a technique called the optical filter method, as schematically shown in Fig. \ref{fbg}.
The technique uses an optical filter to shape the spectrum of the incoming light to FBG with the sample so that the intensity of the reflected light by the FBG is changed with the shift of the Bragg wavelength.
In this way, one can detect the magnetostriction of the sample with the intensity of the light, which can be detected as fast as 100 MHz \cite{IkedaRSI2017} with a resolution of $2\times10^{-5}$.
The technique is applied to various samples, including frustrated spin systems \cite{GenPNAS2023, NomuraNC2023} and even to liquid oxygen \cite{NomuraPRB2021}. 
The method can also be used for millisecond measurement with a better resolution of  $10^{-6}$ below 2 K \cite{IkedaRSI2018}, which has been successfully applied for magnetic insulator volborthite \cite{IkedaPRB2019}, a ternary metallic sample Eu122 system \cite{NakamuraPRB2023}, a $5f$ heavy fermion system UTe$_{2}$ \cite{MiyakeJPSJ2022}, and many other system.

\begin{figure}
\begin{center}
\includegraphics[width = 0.9\columnwidth]{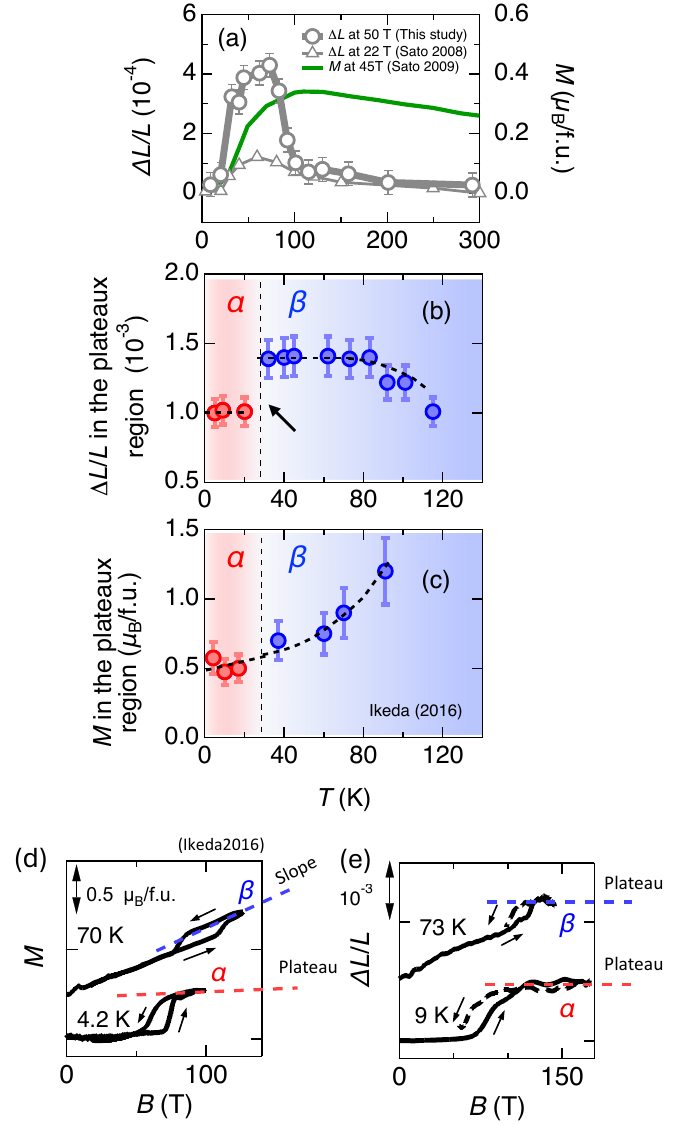}
\caption{(a) Magnetostriction at 20 T and 50 T with magnetization at 45 T as a function of temperature.
(b) Magnetostriction and (c) magnetization in magnetostriction plateau above 100 T as a function of the initial sample temperature.
(d) Magnetization and (e) magnetostriction behaviors at around 70 K and below 10 K shown for comparison.
Reprinted from Ref. \onlinecite{IkedaPRL2020} with a re-arrangement, $\copyright$ 2020 Licensed under CC BY 4.0. \label{fbg02}}
\end{center}
\end{figure}

\begin{figure*}
\begin{center}
\includegraphics[width = 0.8\textwidth]{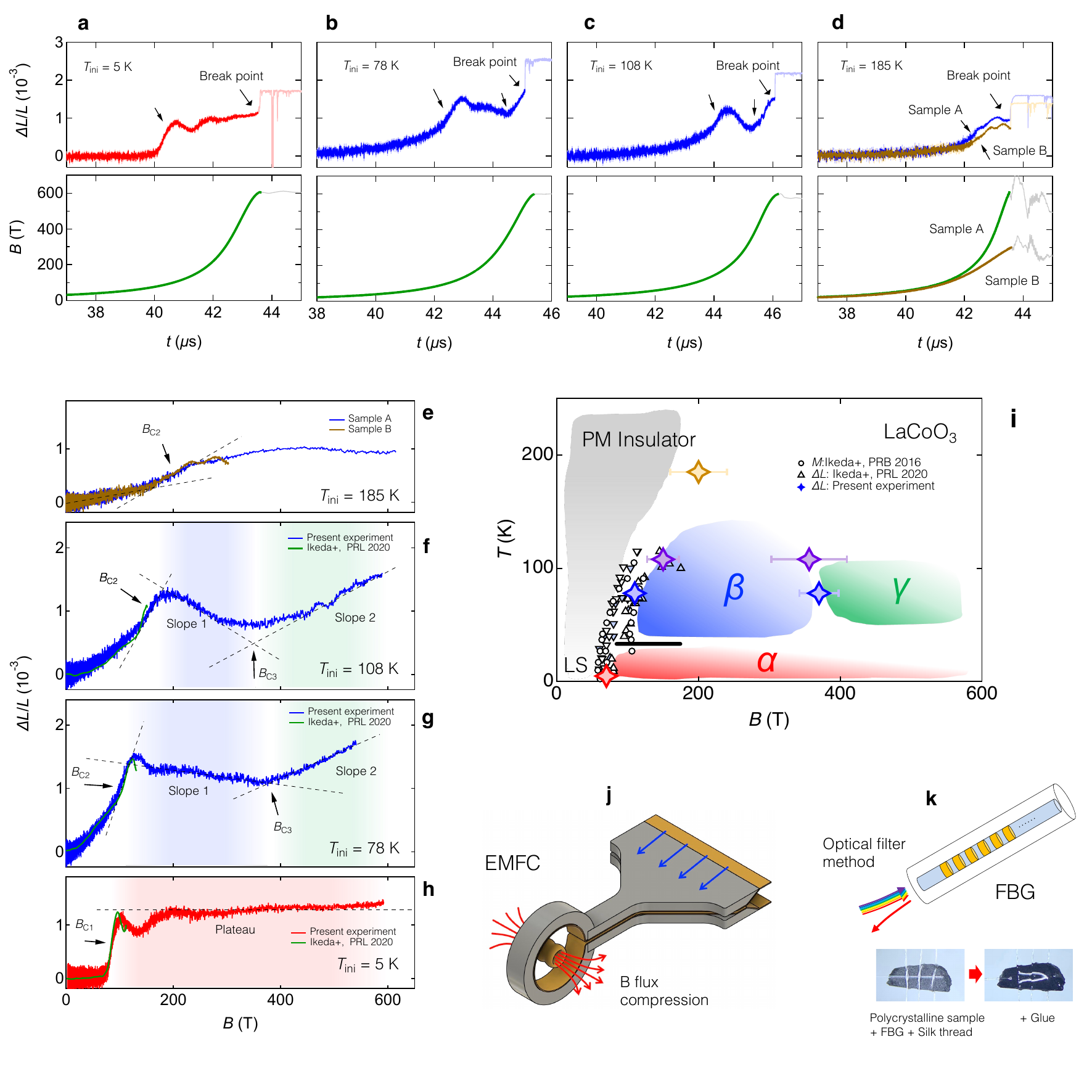}
\caption{Results of magnetostriction measurement on \lco{} up to 600 T using electromagnetic flux compression and FBG with optical filter method. 
Reprinted from Ref. \onlinecite{IkedaNC2023}, $\copyright$ 2023 Licensed under CC BY 4.0. \label{fbg600t}}
\end{center}
\end{figure*}

\subsection{Results}

First, using the magnetization measurements, we have successfully observed the magnetic field-induced phase transition up to 130 T for a collection of non-oriented single crystalline samples \cite{IkedaPRB2016LCO}.
The results are shown in Fig. \ref{mag100t}.
By making the most of the magnetic field range of 130 T, more significant than the previous studies, we have further investigated the temperature dependence of the magnetic field-induced phase transition up to the initial temperature of 120 K.
Surprisingly, as one can see in the $dM/dB$ data as a function of magnetic fields, the transition magnetic field suddenly jumps from 60-70 T to more than 100 T when the initial temperature exceeds 30 K.
Another noteworthy trend is that the transition magnetic field positively depends on the initial temperature beyond 30 K.
The most critical finding is two kinds of high magnetic field-induced phases at magnetic fields beyond 70 T and 100 T with initial temperatures below and above 30 K, respectively, summarized and clearly shown in the phase diagram in Fig. \ref{mag100t}.
This was a counter-intuitive result.
In a simple spin crossover model of a single ion, the magnetic field-induced spin crossover will happen at the lower magnetic field with increasing temperature because spin crossover to the excited state is induced cooperatively by the temperature and magnetic field increases.
Thus, the experimental observation denies the single-ion picture.
Instead, it strongly supports the substantial contribution of inter-site interactions leading to the many-body phases at high magnetic fields.
We here term the high magnetic field phases at low temperature and the high magnetic field phases at high temperature as the $\alpha$ phase and the $\beta$ phase, respectively.
The result does not reproduce the previously observed two-step phase transition at 60 T and 70 T in \onlinecite{MoazPRL2012}.
This may be because we used a collection of non-oriented single crystals in the measurement above 100 T.

To obtain further insights into the observed high magnetic field-induced phases above 100 T, we resorted to magnetostriction measurements up to 150 T with a wide temperature range between 9 - 292 K \cite{IkedaPRL2020}.
As one can see by comparing the magnetization data in Fig. \ref{mag100t} and in magnetostriction data in Fig. \ref{fbg01}, the trend of the transition magnetic fields in the magnetization measurements are reproduced in the magnetostriction measurement, which is further extended to higher magnetic fields up to 150 T.
The striking finding in the magnetostriction data up to 150 T is that magnetostriction $\Delta L/L$ changes at the transition are markedly distinct between the transition to the $\beta$ phase and to the $\alpha$ phase.
This is the height of $\Delta L/L$ at the plateau region in Fig. \ref{fbg01}(c).
They are summarized in Fig. \ref{fbg02}(b).
The distinct features are seen in the magnetostriction data.
In contrast, it is not observable in the magnetization data.
Experimentally, magnetostriction measurements and magnetization measurements are different at several points.
First, the former is a single-shot experiment, whereas the latter requires two shots for each data point, leading to data at higher magnetic fields for magnetostriction measurement.
Secondly, magnetostriction is sensitive to $\Delta L$ itself, whereas the magnetization measurement is sensitive to $dM/dt$.
This fact leads to better sensitivity at the top region of the pulsed magnetic field for magnetostriction measurements than for magnetization measurements.
This fact leads to the present findings made using magnetostriction data.
The result strengthens the evidence that the $\alpha$ and the $\beta$ phases are distinct.

\begin{figure}
\begin{center}
\includegraphics[width = 0.7\columnwidth]{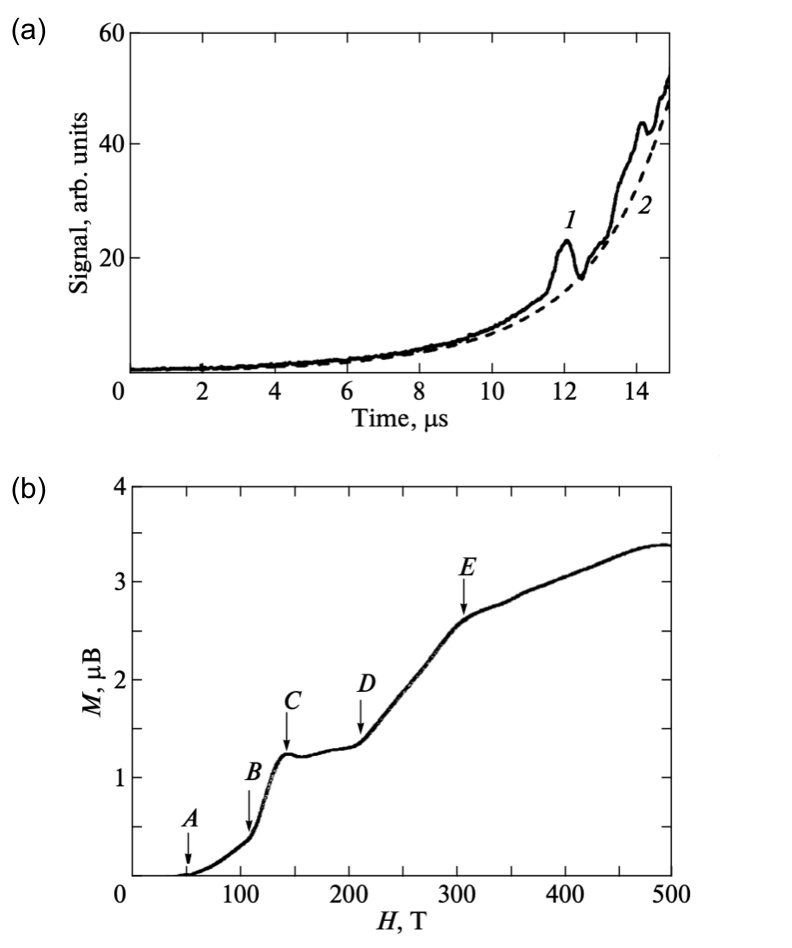}
\caption{Result of a magnetization measurement using induction method and an explosive driven flux compression technique.
Reprinted from Ref. \onlinecite{Platonov2012} with a re-arrangement, $\copyright$ 2012 Springer. \label{platonov}}
\end{center}
\end{figure}

To further extend the phase diagram, we measured magnetostriction on \lco{} up to 600 T, using electromagnetic flux compression and FBG with optical filter method.
We have obtained results with 4 different initial temperatures of 5, 78, 108, and 185 K.
The results are shown in Fig. \ref{fbg600t}a-\ref{fbg600t}d.
The magnetic fields are generated commonly with a maximum magnetic field of about 600 T.
However, the wave profiles are not consistent with each other.
For example, the time to reach the maximum magnetic field varies from shot to shot.
This results from the fact that the total energy of the pulse power varied from shot to shot due to the instability of the air gap switches.
As seen in Fig. \ref{fbg600t}e-\ref{fbg600t}h, the behavior of magnetostriction changes with the changes in the initial temperatures.
At 5 K, magnetostriction shows an abrupt increase, followed by a long plateau feature.
Note that the plateau feature is disturbed by a decaying oscillation, which is a shock wave generated by the abrupt magnetostriction jump, also observed in the previous studies up to 150 T using the single turn coil method \cite{IkedaPRL2020}.
At 78 K, we observed a very sharp feature of slopes with negative and positive slopes.
The change of the sign of the slopes from negative to positive occurs at 380 T.
At 108 K, a similar feature is observed with a more thermally smeared nature.
This feature is more smeared at a higher temperature of 185 K.
We summarize the observed feature in the temperature-magnetic field phase diagram in Fig. \ref{fbg600t}i.
The phase diagram is constructed using the previous results up to 150 T \cite{IkedaPRB2016LCO, IkedaPRL2020} in combination with the present results up to 600 T.

Note that there is a previous report on the magnetization measurement on \lco{} up to 500 T at 4.2 K \cite{Platonov2012}.
The magnetic field was generated using a flux compression method driven by explosives \cite{HerlachRPP1999}.
The magnetization data is measured using a single shot induction method, where the background of the magnetic fields is subtracted, assuming that the background signal is proportional to the generated magnetic fields.
As seen in Fig. \ref{platonov}(a), the raw data of the $dM/dt$ signal shows several peak structures indicating some magnetization increase.
The obtained magnetization data is presented as a function of magnetic fields in Fig. \ref{platonov}(b).
The data does not agree with our experimental data nor the published data below 100 T.
The data may need some calibration regarding the timing and background signal.
However, the experiment makes us believe that the magnetization measurement of \lco{} up to 500 T may be possible with some experimental improvement.

\section{Ideas on the origin of  $\alpha$,  $\beta$, and  $\gamma$ phases}
As presented so far, the magnetic field-induced phases in \lco{} are classified into at least 3 phases: the $\alpha$, the $\beta$, and $\gamma$ phases.
We discuss the possible origins of those phases concerning the electron correlation and spin state degree of freedom in \lco{}.
The spin state degree of freedom is also viewed as the exciton degree of freedom.
Their origins are unclear because experimental evidence in a microscopic viewpoint is missing.
Hopefully, they will be uncovered with a development in microscopic properties at high magnetic fields.

\begin{figure}
\begin{center}
\includegraphics[width = 0.8\columnwidth]{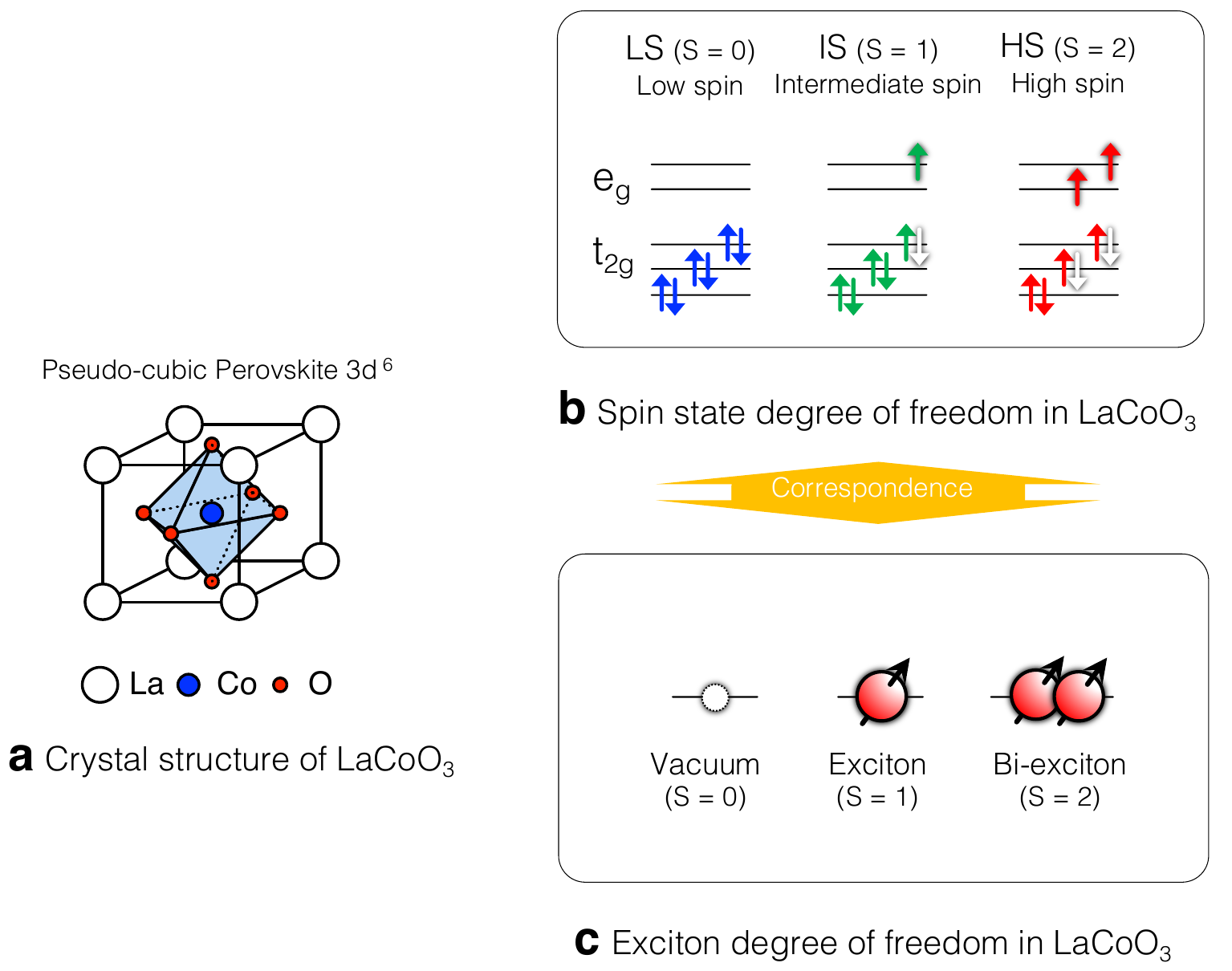}
\caption{Schematic of the perovskite-type structure of \lco{}, the correspondence between the spin states and the exciton state. 
Reprinted from Ref. \onlinecite{IkedaNC2023}, $\copyright$ 2023 Licensed under CC BY 4.0. \label{exciton}}
\end{center}
\end{figure}

\begin{figure}
\begin{center}
\includegraphics[width = 0.9\columnwidth]{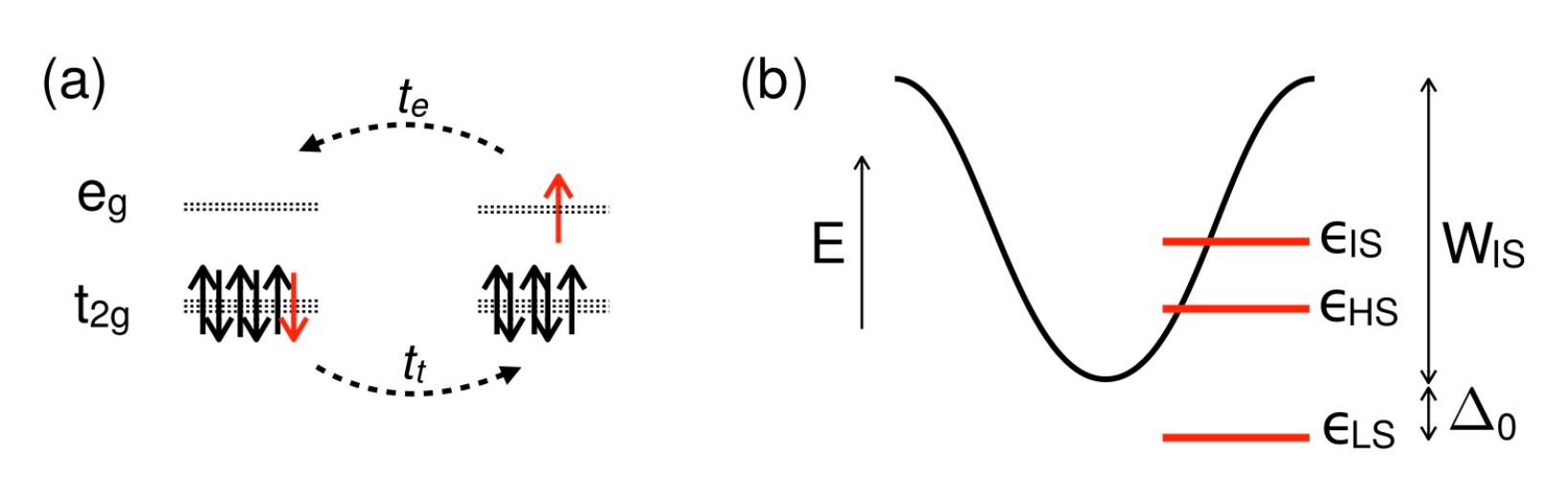}
\caption{Schematic of the effect of dispersion that makes the intermediate spin state ($S=1$) lowest excited state .
Reprinted from Ref. \onlinecite{SotnikovSR2016}, $\copyright$ 2016 Licensed under CC BY 4.0. \label{kunes}}
\end{center}
\end{figure}

\begin{figure}
\begin{center}
\includegraphics[width = 0.9\columnwidth]{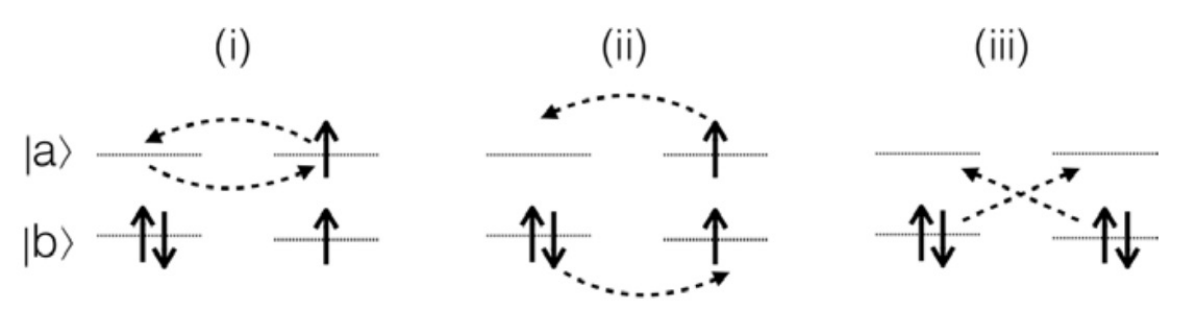}
\caption{Origing of the repulsive interactions between spin triplet excitons (the intermediate spin states $S=1$), and the itineracy of the spin triplet excitons (the intermediate spin states $S=1$).
Reprinted from Ref. \onlinecite{KunesJPCM2015}, $\copyright$ 2012 Institute of Physics. \label{kunes}}
\end{center}
\end{figure}

\subsection{Exciton degree of freedom in \lco{}}
As described in Fig. \ref{exciton}, \lco{} has three possible spin states from high-spin, intermediate-spin, and low-spin states with $S=2$, 1, and 0, respectively.
These states are realized due to the competition between octahedrally coordinated crystal electric field and Hund's coupling in the $3d^{6}$ electrons in for Co$^{3+}$. 
The original Tanabe-Sugano diagram with 3$d^{6}$ system states that either low-spin or high-spin states can be a ground state with the intermediate spin state always being higher in energy than the first excited state.
With an introduction of inter-site interaction, trigonal distortion to the crystal field and spin-orbit couplings can modify the original Tanabe-Sugano diagram, leading to the possible involvement of the intermediate spin state.
This is why the intermediate spin state is considered in addition to the other two states.
Kune\v{s} {\it et al.} primarily focus on the itineracy of the intermediate spin state ($S=1$), leading to the possible lowest excited state \cite{SotnikovSR2016}.
This is schematically shown in Fig. \ref{kunes}.


\subsection{Spin state crystal and exciton condensation}

\begin{figure}
\begin{center}
\includegraphics[width = 0.9\columnwidth]{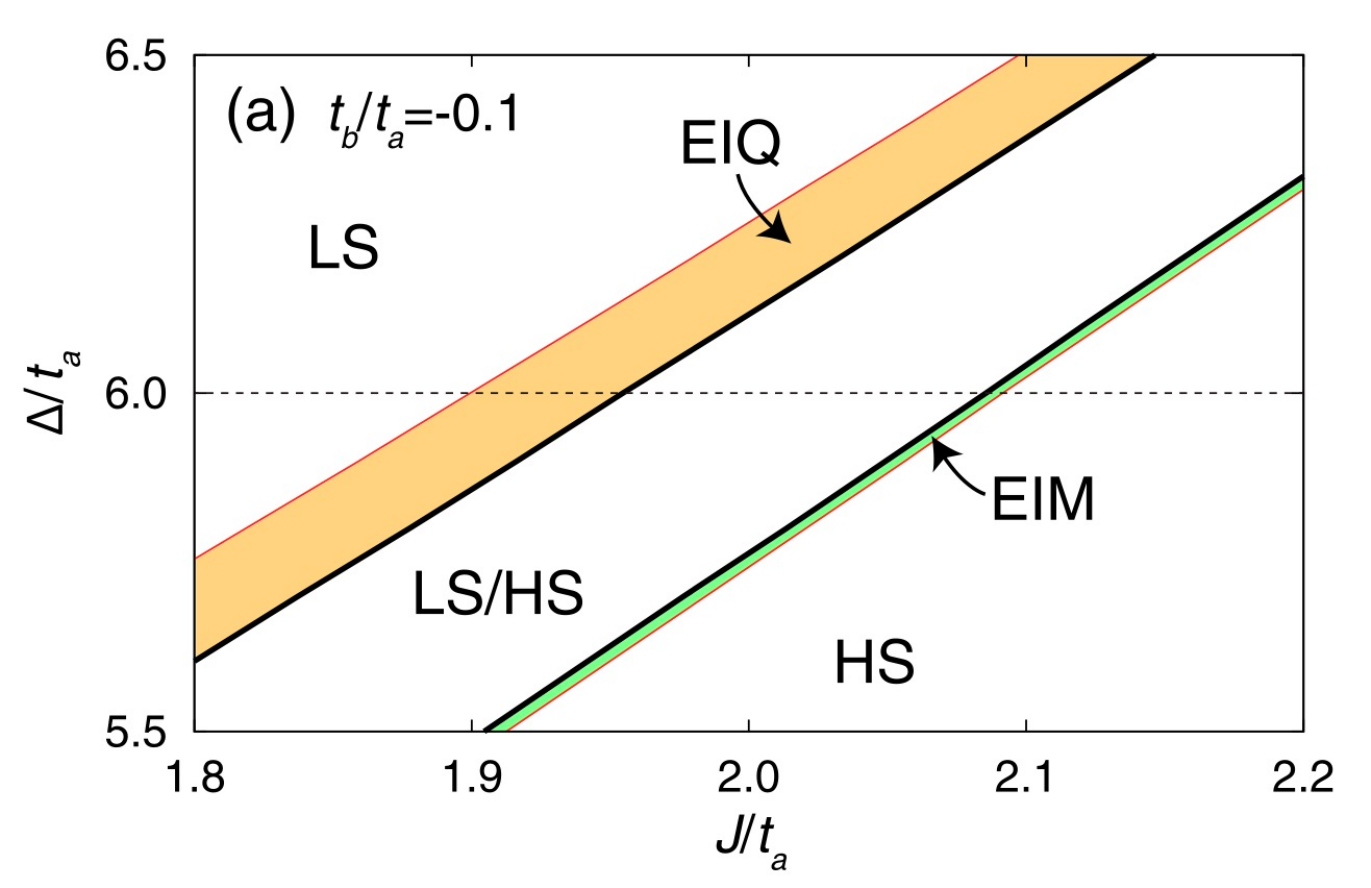}
\caption{Emergence of spin state ordered phase between the uniform low spin state and the uniform HS states.
Here, HS is a $S=1$ state, that can be classified into the intermediate spin state $S=1$ in \lco{}.
Excitonic insulator (EI) phases also appear at the intermediate region.
Reprinted from Ref. \onlinecite{NasuPRB2016}, $\copyright$ 2012 American Physical Society. \label{nasu}}
\end{center}
\end{figure}

\begin{figure}
\begin{center}
\includegraphics[width = 0.9\columnwidth]{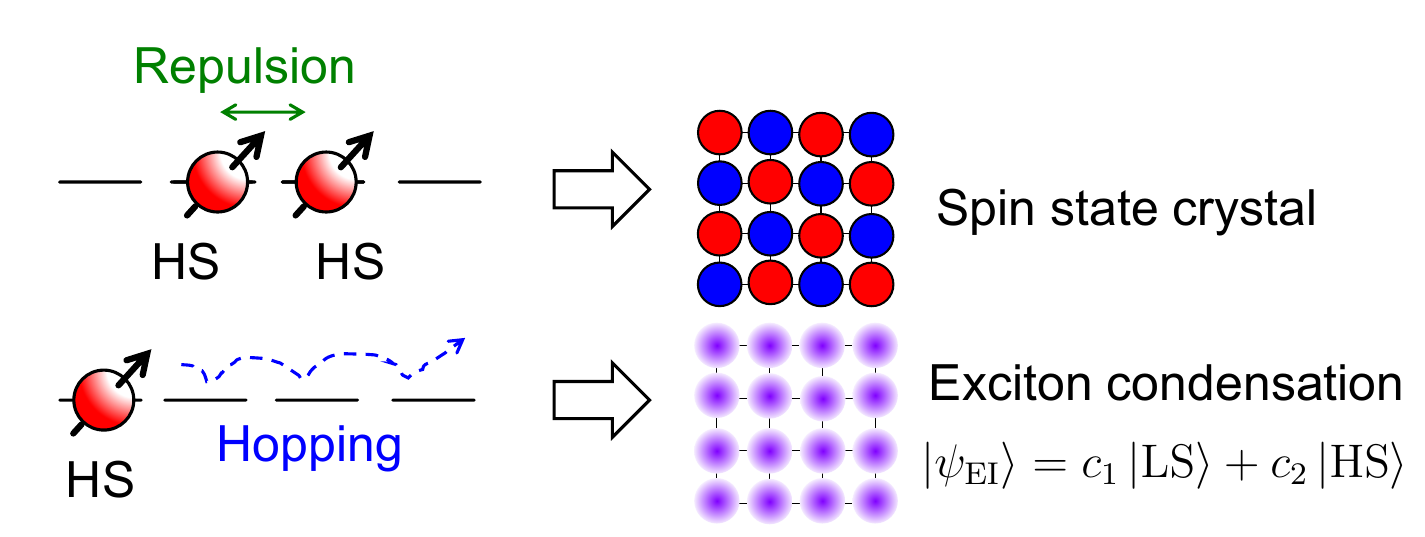}
\caption{The correspondence between spin states and Here, again, HS is a $S=1$ state that can be classified into the intermediate spin state $S=1$ in \lco{}.
\label{fig}}
\end{center}
\end{figure}

\begin{figure}
\begin{center}
\includegraphics[width = 0.9\columnwidth]{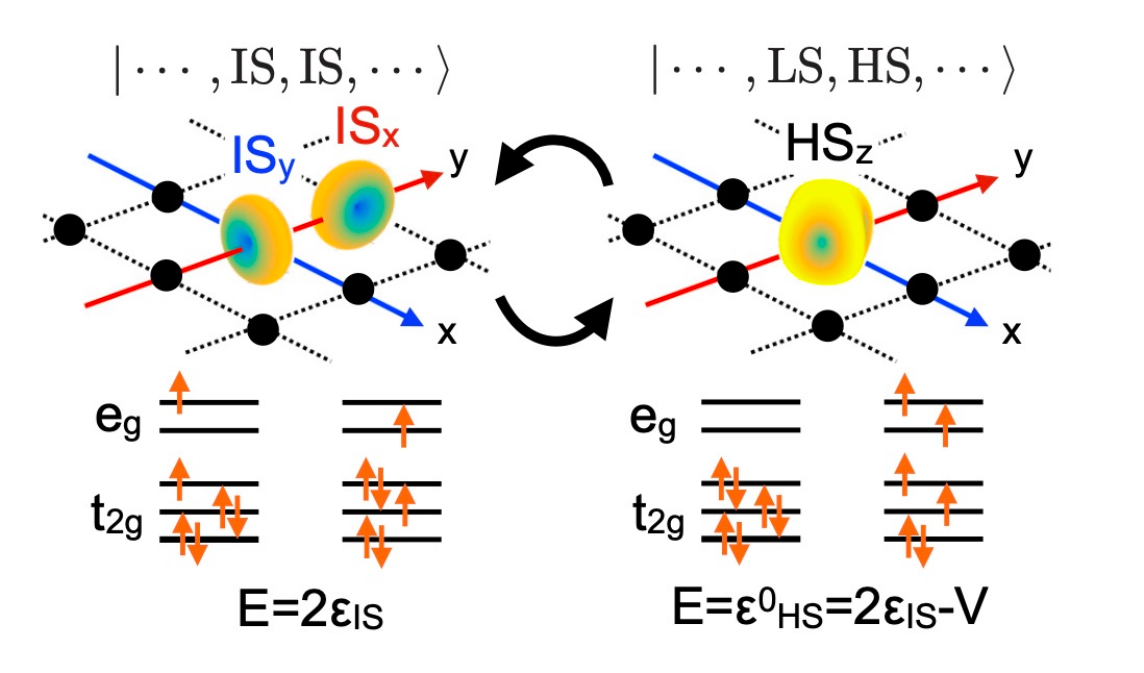}
\caption{Schematic of the hybridization between a pair of the high spin state ($S=2$) and the low spin state, a pair of two neighboring intermediate spin states ($S=1$).
Reprinted from Ref. \onlinecite{HarikiPRB2020}, $\copyright$ 2020 American Physical Society. \label{hariki}}
\end{center}
\end{figure}

Nasu {\ et al.} investigated a ground state phase diagram of the two-orbital Hubbard model by a mean-field analysis \cite{NasuPRB2016}.
The phase diagram is constructed as functions of crystal field splitting $\Delta$ and Hund's coupling $J$ with the electron hopping parameter $t_{a}$ between the $e_{g}$ orbitals and $t_{b}$ between the $t_{2g}$ orbitals.
At the limit of large $J$, it is trivial that a polarization to the HS occurs.
At the limit of large $\Delta$, it is again natural to obtain the low spin phase.
With an introduction of inter-site interactions $t_{a}$ and $t_{b}$, non-trivial phases such as spin state crystals and exciton condensation emerge.
These two phases are incompatible and competing with each other.
A spin state ordered phase of the intermediate spin state and low spin state emerges between the spatially uniform low spin state and the spatially uniform intermediate spin state as shown in Fig. \ref{nasu}.
in Fig. \ref{nasu}, they term the intermediate spin state $S=1$ to be a HS state.
Microscopically, spin state crystallization occurs due to the repulsive interactions between the intermediate spin states, which is, in other words, the attractive interaction between the low spin state and the intermediate spin state as is shown in Fig. \ref{fig}.
In addition to spin state crystal phase, the excitonic insulator phase appears at the boundary between the spin state crystal and low spin state, and the boundary between the spin state crystal and the high spin state.
Due to its itineracy, this phase appears at low temperatures as a Bose-Einstein condensation of the spin triplet exciton.
The occurrence of exciton condensation in the spin crossover systems is also reported in literature \cite{KanekoPRB2012, KunesPRB2014, KunesJPCM2015, KanekoPRB2015}.
The occurrence of spin state crystal and exciton condensation is schematically depicted in Fig. \ref{fig}.

\begin{figure}
\begin{center}
\includegraphics[width = \columnwidth]{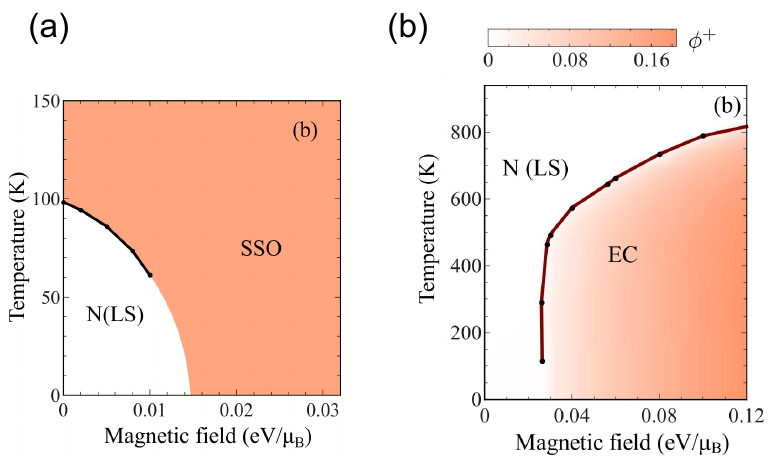}
\caption{A calculated temperature-magnetic field phase diagrams of \lco{} with low spin ground state and (a) spin state crystal and 
 (b) exciton condensation
 as the excited state.
Reprinted from Ref. \onlinecite{SotnikovSR2016} with a re-arrangement, $\copyright$2016 Licensed under CC BY 4.0. 
\label{sotnikov2}}
\end{center}
\end{figure}

\begin{figure}
\begin{center}
\includegraphics[width = 0.9\columnwidth]{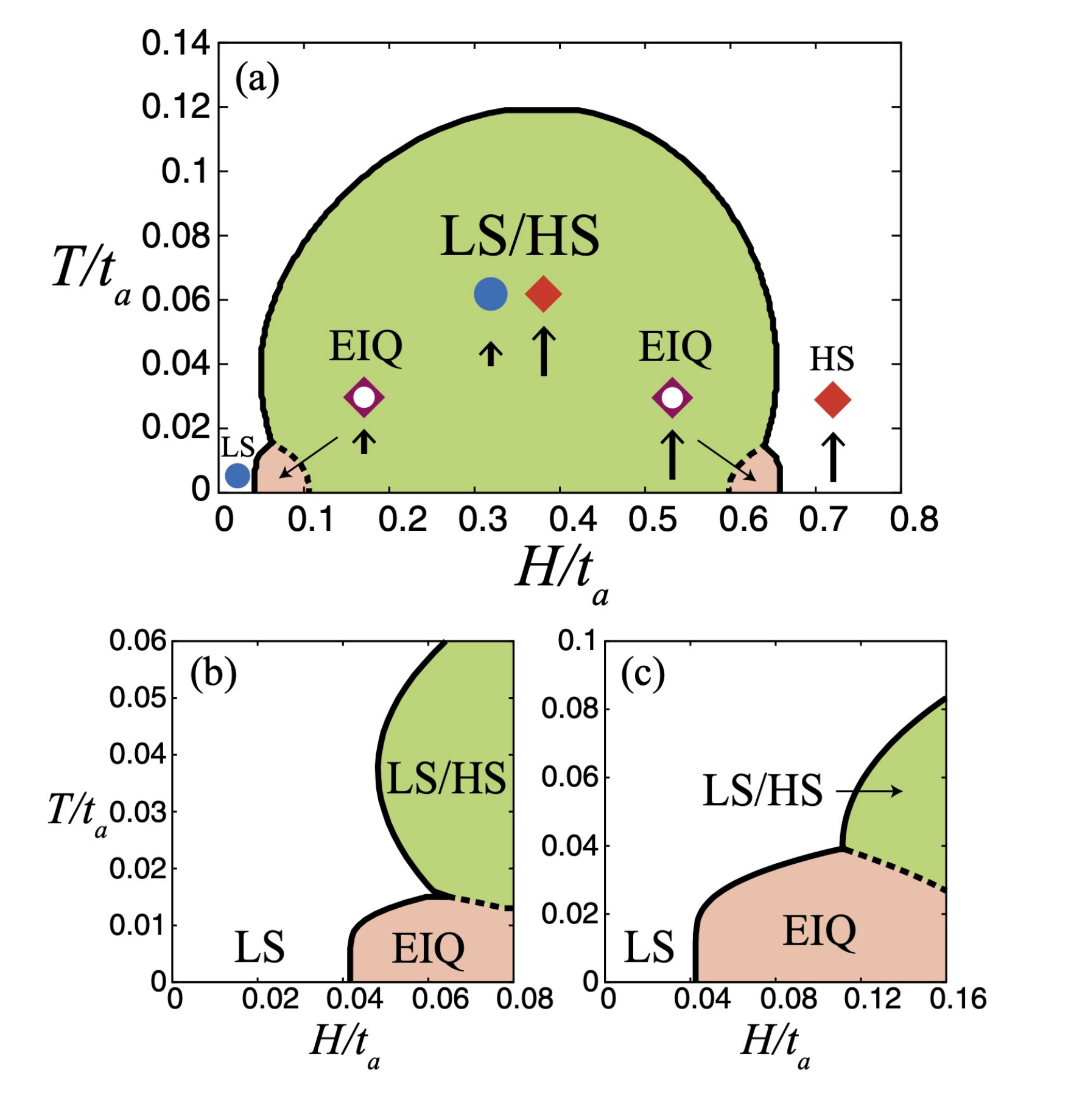}
\caption{(a) A calculated temperature-magnetic field phase diagram for \lco{}. It is calculated using a two-orbital Hubbard model and a mean-field approximation.
(b) A closed-up of a region occurs at the first magnetic field-induced spin crossover.
(c) same area with slightly different parameters.
Reprinted from Ref. \onlinecite{TatsunoJPSJ2016}, $\copyright$ 2016 The Physical Society of Japan.
\label{tatsuno}}
\end{center}
\end{figure}

\begin{figure}
\begin{center}
\includegraphics[width = 0.8\columnwidth]{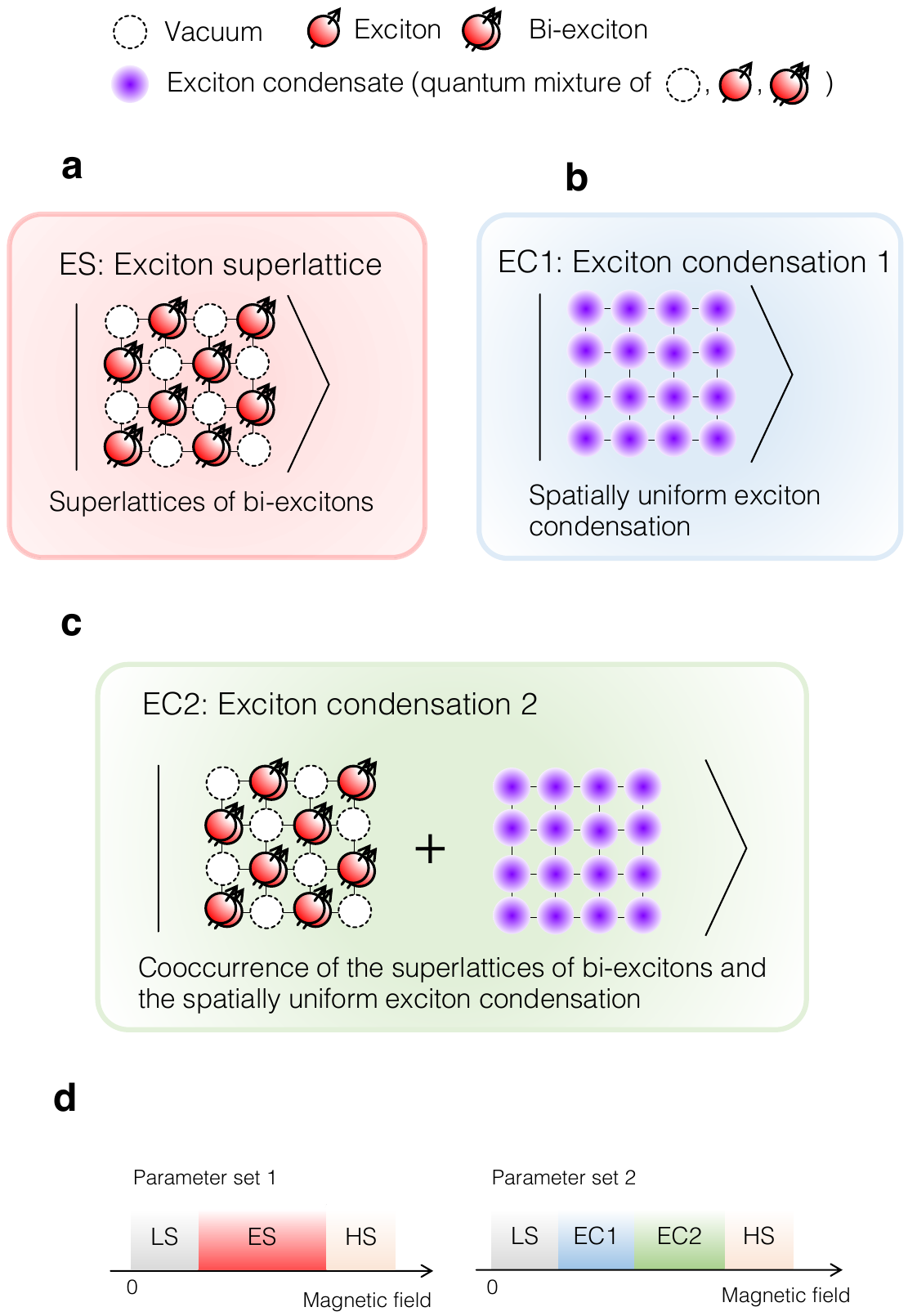}
\caption{Schematic drawing of (a) spin state crystal of high spin state $S=2$ and low spin state $S=0$ (b) Exciton condensation (c) Co-occurrence of the spin state crystal and exciton condensation (d) a phase diagram for 5 K in magnetic fields with a parameter set 1 (e) a phase diagram for 78 K in magnetic fields with a parameter set 2.
Reprinted from Ref. \onlinecite{IkedaNC2023}, $\copyright$ 2023 Licensed under CC BY 4.0. 
\label{600theory}}
\end{center}
\end{figure}

\subsection{Duality of high spin and intermediate spin states}
Two recent studies pointed out an interesting idea that might solve the long-standing controversy over the high spin state and the intermediate spin states.
The high spin state and intermediate spin state can be hybridized if we consider an electron hopping process.
Hariki {\it et al.} suggested this idea \cite{HarikiPRB2020}, which is schematically shown in Fig. \ref{hariki}.
As can be seen, with two electrons hopping between adjacent Co ions, a pair of intermediate spin states can be transformed into a pair of high-spin states and low-spin states.
In more detail, an electron hopping at between $e_{g}$ orbitals of adjacent sites and an electron hopping between $t_{2g}$ orbitals of adjacent sites will realize this change of states.
It is shown that high spin and intermediate spin states can be hybridized.
Such hybridization is supported by the inelastic neutron scattering experiment where a ferromagnetic 7-site cluster has been identified in the magnetic scattering component \cite{TomiyasuArxiv}.

\subsection{Magnetic field induced spin state order}

Sotnikov {\it et al.} and Tatsuno {\it et al.}  independently reported studies on exciton condensation and spin state crystallization induced by high magnetic fields in \lco{} \cite{TatsunoJPSJ2016, SotnikovSR2016}.
They studied the emergence of spin state crystals and excitonic condensation on the temperature-magnetic field plane.
Sotnikov {\it et al.} argued that excitonic condensation is more probable than spin state crystal as the possible origin of $\alpha$ phase.
This is based on the calculated temperature-magnetic field phase diagram that the phase boundary between the low spin state and the field-induced phase has a positive dependence on temperature in the case of exciton condensation and negative dependence on temperature in the case of spin state crystal, as shown in Fig. \ref{sotnikov2}.
Tatsuno {\it et al.} showed that spin state crystal and excitonic condensation are induced at high magnetic fields with high and low temperatures, respectively, as shown in Fig. \ref{tatsuno}.
They indicate that the $\alpha$ and $\beta$ phases can be attributed to exciton condensation and spin state crystal.
It was argued that, with extensive magnetic fields, the re-entrant to exciton condensation occurs before the complete saturation of the magnetic moment.

\begin{figure}
\begin{center}
\includegraphics[width = 0.9\columnwidth]{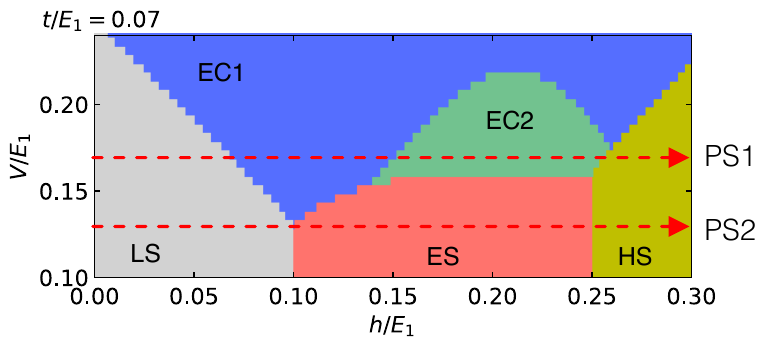}
\caption{Phase diagram the effetive model of \lco{} on the parameter space of the hybiridization of high spin state and the intermediate spin state $V$, and the magnetic field $h$ devided by the crystal field splitting $E_{1}$ with a hopping parameter $t$ of the intermediate spin state.
Reprinted from Ref. \onlinecite{IkedaNC2023}, $\copyright$ 2023 Licensed under CC BY 4.0. 
\label{params}}
\end{center}
\end{figure}

Motivated by the experimental results up to 600 T, further calculation has been performed \cite{IkedaNC2023}.
Theoretical calculations so far considered only the intermediate spin state as an excited spin state.
The present calculations consider the high spin state of $S=2$, in addition to the low spin $S=0$ and the intermediate spin state $S=1$.
In addition to the interaction considered in previous calculations, the hybridization between high spin and intermediate spin states is taken into account \cite{IkedaNC2023}.
The calculated result shows a variety of excitonic condensation and spin state crystallization, and their co-occurrence has been investigated as shown in Fig. \ref{600theory}.
We found a parameter set 1 (PS1) and a parameter set 2 (PS2) that reproduce the appearance of two kinds of exciton condensation and spin state crystallization, respectively, as shown in Fig. \ref{params}.
This suggests that the interaction parameters drastically changed between the temperatures of 5 K and 78 K.
Such lattice change between $\alpha$ and $\beta$ phases are confirmed in the previous study \cite{IkedaPRL2020} as shown in Fig. \ref{fbg02}.



\section{Summary}
\subsection{Conclusion}
We have reviewed experimental studies on \lco{} at high magnetic fields.
Experimentally, several novel phases, $\alpha$, $\beta$, and $\gamma$ phases have been discovered. 
The appearance of multiple phases strongly suggests that inter-site interactions between spin states are in play, such as the itineracy of the intermediate spin state and the hybridization (duality) of the intermediate spin state and the high spin state.
However, the experimental evidence is limited to some macroscopic studies, and a concrete possibility of the origin of those phases has not yet been obtained.
We have introduced theoretical studies on \lco{} at high magnetic fields.
Those studies focus on the Mott-Hubbard models with two orbitals, a minimal model for correlated electron systems with spin state degrees of freedom.
Various ordered phases, including exciton condensation and spin state crystals, are identified.

\subsection{Future prospects}
Further exploration on \lco{} is needed to clarify the origin of $\alpha$, $\beta$, and $\gamma$ phases using various experimental methods available at high magnetic fields such as dielectric constant, ultrasound, electric conductivity.
A microscopic experiment in high magnetic fields, such as x-ray diffraction and terahertz spectroscopy, is even more critical.
X-ray techniques at 100 T are being developed using x-ray free electron laser and portable 100 T generators called PINK series \cite{IkedaAPL2022, IkedaPRR2020}.
Single-shot Teraheltz time domain spectroscopy is reported, which is also a promising application with the PINK series for a 100 T experiment.
Future studies will uncover the origin of the high-field phases, leading to a more profound understanding of the correlation effect in systems with a spin state degree of freedom.

\begin{acknowledgements}
This work is supported by MEXT LEADER program No. JPMXS0320210021, JST FOREST program No. JPMJFR222W, JSPS Grants-in-Aid for Scientific Research No. 23H04861, 23H04859, 23H01121.
\end{acknowledgements}
\bibliography{jpsj}
\end{document}